\documentclass[a4paper,11pt]{article}

\usepackage{jcappub} 

\usepackage[T1]{fontenc} 
\usepackage[utf8]{inputenc}

\usepackage{natbib}

\usepackage{graphicx}
\usepackage{dcolumn}
\usepackage{bm}

\usepackage{array}
\usepackage{booktabs}
\usepackage{tabu}
\usepackage{dcolumn}
\usepackage{multirow} 
\usepackage{amsmath}
\usepackage{amsfonts}
\usepackage{amssymb}
\usepackage{graphicx}
\usepackage{subfigure}
\usepackage{graphicx}
\usepackage{dcolumn}
\usepackage{bm}
\usepackage{xcolor}

\graphicspath{{figs/}} 

\title{\boldmath Effects of the inflaton mass on the pre-inflationary dynamics and primordial power spectra in loop quantum cosmology}

\author[]{Kui Xiao}
\author[1]{Abolhassan Mohammadi\note{Corresponding author.}}
\author[]{Hongwei Tan}

\affiliation[]{School of Science, Hunan Institute of Technology, Hengyang 421002, China.}

\emailAdd{87xiaokui@gmail.com}
\emailAdd{abolhassanm@hnit.edu.cn}
\emailAdd{honweitan@hnit.edu.cn}

\abstract{Considering the mass parameters as free phenomenological parameters, we investigate inflation driven by a canonical scalar field with different potentials in LQC. We find that a smaller mass value makes it easier to obtain a sufficient number of e-folds for the quadratic potential, while for the Starobinsky potential, more e-folds are obtained for larger mass values. By estimating the probability of slow-roll inflation, we find that it remains very close to unity for all considered cases. We then compute the primordial power spectrum using three different approaches as the dressed metric, the hybrid, and the alternative mass function approaches for the case that the kinetic energy dominated at bounce. The resulting power spectrum for the both potentials and for different mass values shows the same qualitative pattern in all three approaches. It is suppressed at small $k$, amplified and oscillating over an intermediate range, and nearly scale-invariant at large $k$. The results from approaches are mainly differ in how fast the power spectrum converges to this scale-invariant regime, determining that the hybrid approach converging the fastest, followed by the alternative mass function approach, and then the dressed metric approach. Then, we determine the pivot scale $k_\star$ and finds that it is highly sensitive to the inflaton mass, changing by more than an order of magnitude for a change in mass of only a few percent. Using the numerically obtained tensor power spectrum at the pivot scale $k_\star$, we calculate the scalar spectral index $n_s$ and the tensor-to-scalar ratio $r$. The resulting values of $n_s$ and $r$ are found to be in good agreement with the current data. The resulting scalar power spectra are finally fed into the CAMB code to obtain the corresponding angular power spectrum, and compared with the Planck 2018 data and the best-fit $\Lambda$CDM model. All three approaches shows a consistency with data at high multipoles, while at the low multipoles the  hybrid approach gives the closest agreement with the data, and the dressed metric approach shows the largest deviation from the best-fit $\Lambda$CDM curve.}

\begin{document}
\maketitle
\flushbottom

\section{Introduction}\label{sec1}
Inflation theory was introduced to solve several cosmological fine-tuning problems, especially the horizon, homogeneity, and flatness problems \cite{inflationtheory1,inflationtheory2,inflationtheory3,inflationtheory4}. For recent reviews of inflation theory, see \cite{Gonzalez}. In addition to solving these problems, inflation provides a natural mechanism for generating primordial density perturbations from quantum fluctuations, which later evolve into the observed large-scale structures of the Universe \cite{Lyth,Planck2020}. Based on the inflationary scenario, a quantum field evolves on a classical spacetime background described by general relativity (GR) \cite{Lyth,Calcagni}. In the simplest inflationary models, the potential energy of the inflaton dominates over its kinetic energy during the slow-roll phase, leading to an approximately constant Hubble parameter. Current observations indicate that inflation should last at least about $55-65$ e-folds. However, the initial conditions of inflation remain an open question \cite{Brandenberger}, since direct observational information about the very early Universe is still unavailable. Moreover, within the framework of classical GR, the existence of a big-bang singularity is unavoidable when the matter content satisfies the standard energy conditions. This singularity problem may indicate the limitation of Einstein's theory at extremely high energy scales.

To solve the big bang singularity problem, one possible solution is to modify the theory of general relativity at high energy scales. There are many candidates, one of which is loop quantum cosmology (LQC; for more recent reviews, please see \cite{Calcagni,Bojowald-Book,AgulloReview,BarrauReview}). LQC is a canonical quantization of homogeneous spacetimes based on the techniques used in loop quantum gravity \cite{Rovelli-Book,Thiemann-Book}. The dynamics of LQC can be studied effectively by introducing quantum gravity corrections to the gravitational and matter Hamiltonians \cite{Bojowald-Book}, and evidence shows that the effective equations provide an excellent approximation to the full dynamics of sharply peaked states \cite{Ashtekar-74,Corichi,Rovelli,Diener,Diener-1}. In the LQC scenario, one can obtain many interesting results, e.g., the replacement of the big bang by the big bounce \cite{Ashtekar-74,Ashtekar,Ashtekar2}, the avoidance of most singularities \cite{Singh,Vereshchagin,GDate,PSingh,PSingh2,PSingh3,Kauzuharu,Joe}, the more likely occurrence of inflation \cite{Singh-in,Corichi2,Sloan,Sloan2,Singh-in-2,Linsefors,Chen,Bedic}, and so on. Note that there are some new developments in LQC, for example, the phenomenon of signature change \cite{BojowaldSignature1,BojowaldSignature2,BojowaldSignature3}, and various possible loop quantizations of cosmological spacetimes in the framework of LQG \cite{Yang,Li084029}.

To discuss the evolutionary pictures of the Universe, one always set the initial condition at the bounce point. There is a desired slow-roll inflation while the evolution is initially dominated by the kinetic energy of the scalar field of at the quantum bounce \cite{Ashtekar-report,Singh-in,Mielczarek,Zhang,BollietMass, Bonga1,Bonga2}. Considering the pre-inflation dynamical behavior, the slow-roll inflation of canonical scalar field with different potentials have been investigated \cite{Corichi2,Sloan,Sloan2,Bonga1,Bonga2, Barrau,powerlaw,zhu-preinflation,zhu-preinflation2,Shahalam2,Universe,Sharma,inflation}. Inflation with a non-canonical scalar field has also been discussed \cite{Zhang2,SenPRD,Xiong,Xiao,XiaoTach,Mohammadi_DBI}. As same as the background evolution, the perturbations of inflation in LQC scenario have been investigated. Up to now, there are at least four types of perturbation theories to study the primordial power spectra in LQC scenario, i.e., hybrid approach \cite{MaruganHA1,MaruganHA2,MaruganHA3,MaruganHA4}, the dressed metric approach \cite{AgulloDM1,AgulloDM2,AgulloDM3}, deformed algebra approach \cite{BojowaldCD1,BojowaldCD2,BojowaldCD3}, separate universe \cite{EwingSU1,EwingSU2}, and the recently proposed alternative mass function \cite{LiSingh2024mass,MohammadiLiZhu2025}. Among these, the dressed metric and the hybrid approaches are built based on the same foundation so that the background is loop quantized and the perturbations are Fock quantized; however, they result in different effective masses once the classical mass is polymerized \cite{Navascues,ElizagaNavascues:2018bgp,LiSingh2022}. The alternative mass function approach is obtained from the classical mass function written in the comoving gauge of the Lagrangian formulation of perturbation theory, leading to a distinct effective mass function \cite{LiSingh2024mass}. In contrast to the dressed metric and the hybrid approaches, the effective mass in the alternative mass function approach contains a correction term that cannot be neglected near the bounce, even in the case of a kinetic-dominated bounce. Considering the primordial power spectrum and the corresponding angular power spectrum shows that the three approaches agree closely with the observed spectrum at high multipoles, while the differences between them become visible only at the lowest multipoles, where the imprint of the quantum bounce is expected to be strongest \cite{LiWangSingh2020DM,LiOlmedoSingWang2020Hyb,MohammadiLiZhu2025}.

In order to compare the relationship between theoretical analysis results and observed data, the mass of inflaton is chosen to be in an agreement with the observational data (i.e., see appendix of \cite{Shahalam2}). In this case, the slow-roll inflation always happens in GR dominated stage, and the quantum effect can be ignored. Considering $m$ is a variable, Ref.\cite{BollietMass} discussed the tensor perturbations for three different masses: $m=10^{-3}m_{Pl}$, $m=10^{-2.5}m_{Pl}$ and $m=10^{-2}m_{Pl}$, one found that the trends of primordial power spectra for tensor modes depend on the value of the mass. For the modified LQC, one found that the main properties remain the same for a different choice of m for $m\in(10^{-4},10^{-7})$ \cite{Li066016}. In LQC scenario, whether the value of mass has an impact on the evolution of the universe and the primary energy spectrum is still worth further discussion. 
In this work, considering $m$ as a parameter, we investigate the inflationary dynamics in LQC scenario for the canonical scalar field with two different potentials, the quadratic and the Starobinsky potentials. We further compute the resulting scalar primordial power spectrum using three different approaches to cosmological perturbations in LQC, namely the dressed metric, the hybrid, and the alternative mass function approaches, and study how the inflaton mass affects the power spectrum and the pivot scale in each case. Finally, we compute the corresponding angular power spectrum and compare it with the Planck 2018 data and the best-fit $\Lambda$CDM model.

This work is organized as follows. In Sec. \ref{Sec2}, we briefly introduce the slow-roll inflation theory in LQC. In Sec. \ref{Sec3}, we examine the duration of the slow-roll inflation for two different potentials. This section analyzes in detail the evolution of the background in the framework of four different inflaton masses. 
Then, in Sec. \ref{Sec4}, we numerically estimate the power spectrum using three approaches as dressed metric, hybrid, and alternative mass function, and consider the effect of mass on the resulting power spectrum. The main conclusions are summarized in the last section.

\section{Inflation theory in LQC}\label{Sec2}

In this section, we briefly review the inflation theory in holonomy-corrected LQC. In this work, the inverse-volume corrections are not considered since these corrections have not yet been fully realized in minisuperspace models \cite{Bojowald}. Based on the effective Hamiltonian constraint and the corresponding Hamilton equations in LQC, the modified Friedmann and Raychaudhuri equations are given by
\begin{eqnarray}
	H^2&=&\frac{8\pi}{3 m_{Pl}^2}\rho\left(1-\frac{\rho}{\rho_c}\right)\label{Fri},\\
	\dot{H}&=&-\frac{4\pi}{m_{Pl}^2}(\rho+P)\left(1-\frac{2\rho} {\rho_c}\right)\label{dH},
\end{eqnarray}
where $\rho$ and $P$ are the energy density and pressure of the matter field, respectively, and $m_{Pl}=1/\sqrt{G}$ is the Planck mass. The parameter $\rho_c$ represents the critical energy density and is given by $\rho_c\equiv 3/(8\pi G\lambda^2\gamma^2)\simeq 0.41 \rho_{Pl}$, where $\gamma\simeq0.2375$ is the Barbero-Immirzi parameter \cite{BIP}, and $\lambda^2\equiv4\sqrt{3}\pi\gamma\ell^2_{Pl}$. Here, $H$ is the Hubble parameter defined as $H\equiv\dot{a}/a$, where $a$ is the scale factor and the dot denotes the derivative with respect to the cosmic time. 

It should be noted that the effective equations (\ref{Fri}) and (\ref{dH}) do not include the isotropic reductions of higher-curvature terms. Since LQC models are not derived from a covariant quantum gravity theory, the corresponding higher-curvature corrections have not been fully determined yet \cite{Bojowald00044}. In this work, we consider the matter content of the Universe to be a single massive scalar field $\phi$. The equation of motion for the scalar field is described by the Klein-Gordon equation
\begin{eqnarray}
	\ddot{\phi} + 3H\dot{\phi}+ V_{,\phi} =0 \label{kG},
\end{eqnarray}
where $V(\phi)$ is the scalar field potential and $V_{,\phi}\equiv dV/d\phi$. Inflation driven by a scalar field has been extensively studied in LQC for various types of potentials, including power-law potentials \cite{Sloan,Sloan2,Barrau,powerlaw,zhu-preinflation,zhu-preinflation2,Universe,inflation,Luc}, Starobinsky potentials \cite{zhu-preinflation,Bonga1,Bonga2,Shahalam2,inflation}, monodromy potentials \cite{Sharma}, and other forms of inflationary potentials \cite{Shahalam2}. In this work, we focus on the quadratic potential
\begin{eqnarray}
	V(\phi)=\frac12m^2\phi^2,
\end{eqnarray}
and Starobinsky potential
\begin{eqnarray}
	V(\phi)=\frac{3}{32\pi}M^2m^2_{Pl}\left(1 -e^{-\sqrt{\frac{16\pi}{3}}\frac{\phi}{m_{Pl}}}\right)^2,
	\label{Staro}
\end{eqnarray}
where the parameters $m$ and $M$ have dimensions of mass. It should be noted that considering $R^2$-inflation in LQC results in a different potential from Eq. (\ref{Staro}) \cite{zhangPRL}. For the quadratic potential, the mass parameter $m$ is usually treated either as a constant fixed by observational data \cite{zhu-preinflation,Sloan} or as a free parameter in the phenomenological analysis \cite{BollietMass,Li066016}. Considering the slow-roll approximation, the values of $m$ and $M$ can be constrained from observations. For example, the best-fit results from the Planck 2015 data give $m\simeq 1.26\times10^{-6}m_{Pl}$ \cite{zhu-preinflation} and $M=2.5\times10^{-6}m_{Pl}$ \cite{Bonga1}. Its value can be different in other inflationary models; for example, in the induced gravity scenario, one obtains $M\simeq1.25\times10^{-5}m_{Pl}$ \cite{Pallis}. In this work, we treat both $m$ and $M$ as phenomenological parameters. The Starobinsky mass parameter $M$ can also be interpreted as the mass scale of the inflationary potential around the vacuum state \cite{Ellis,Asaka}.

The modified Friedmann equation (\ref{Fri}) shows that the Hubble parameter vanishes when the energy density reaches its maximum value $\rho_c$. This point is called the bounce point, where the contracting phase of the Universe changes into the expanding phase. After the bounce, the Universe enters a super-inflationary phase until the energy density decreases to $\rho=\frac12\rho_c$. During this period, both the Hubble parameter $H$ and the scale factor $a$ increase.

Following the bounce, the Universe can enter a slow-roll inflationary phase \cite{Sloan,Sloan2,Corichi,Linsefors,Chen,Bedic}. Since the pre-inflationary evolution starts from the quantum bounce, the initial conditions are set at the bounce point. The dynamics with potential-energy-dominated (PED) and kinetic-energy-dominated (KED) initial conditions at the bounce have been widely investigated in the literature \cite{Sloan,Sloan2,Bonga1,Bonga2,Barrau,inflation,powerlaw,zhu-preinflation,Shahalam2,Sharma,Luc}. The total number of inflationary e-folds $N$ depends on the choice of the initial conditions. For the PED case, a very large value of $N$ can be obtained for power-law potentials \cite{powerlaw,zhu-preinflation,Universe,inflation}, while the number of e-folds is usually too small for the Starobinsky potential \cite{Bonga1,Bonga2,zhu-preinflation,inflation}. Similar behavior has also been observed for tachyonic inflation in the LQC framework \cite{XiaoTach}. On the other hand, for the KED case, both the power-law and Starobinsky potentials can naturally produce a suitable number of e-folds, typically around $N\sim60$ \cite{Bonga1,Bonga2,powerlaw,zhu-preinflation,Universe,inflation}.

In the KED case, the evolution of the Universe before the preheating can be divided into three different phases as super-inflation, damping, and slow-roll inflation\footnote{The super-inflationary stage occurs immediately after the bounce and is part of the quantum bounce regime, while the damping stage corresponds to the transition period between the super-inflationary and slow-roll inflationary phases.} \cite{zhu-preinflation,inflation}. During the super-inflation phase, the evolution of the scale factor $a(t)$ is independent of the choice of the potential $V(\phi)$, and one can get an analytical solution $a(t)=$ $a_B\left(1+\gamma_B t^2/t_{Pl}^2\right)^{1/6}$, with $\gamma_B=24 \pi \rho_c$, and the subscript ``B" denotes the value of a variable at the bounce point \cite{zhu-preinflation,zhu-preinflation2,inflation}. During the damping phase, the equation of state gradually evolves from the kinetic-energy-dominated regime with $w\simeq1$ to the potential-energy-dominated regime with $w\simeq-1$, allowing the Universe to enter the slow-roll inflationary phase.

The scalar and tensor perturbations for the KED case during the three phases have been studied \cite{zhu-preinflation}. Just as the evolution of the background, the perturbations are independent of the potential during the super-inflation and damping phase while one considers the dressed metric approach. At the slow-roll inflation phase, the scalar curvature and tensor power spectrum in LQC scenario get modified with respect to GR scenario, such that there can be written as $\mathcal{P}_{S,T}(k)=\left(1+\delta_{Pl}\right) \mathcal{P}^{GR}_{S,T}(k)$, with $S$ for scalar perturbation and $T$ for tensor perturbation, but the tensor-scalar ratio $r$ is the same as that given in GR \cite{zhu-preinflation}.

In most studies, the critical energy density $\rho_c$ and the mass parameter of the scalar field are treated as constants. In the KED scenario, the resulting scalar and tensor power spectra are sensitive to the value of the inflaton mass parameter \cite{BollietMass}. The characteristic scales separating the infrared, oscillatory, and nearly scale-invariant regimes are determined by the background evolution and therefore depend on the parameters of the model, including the critical energy density $\rho_c$. In addition, the mass of the field may affect the probability of the desired slow-roll inflation. As an example, we calculate the probability of slow-roll inflation for the quadratic potential. According to \cite{Sloan2,Graef}, the probability of slow-roll inflation is given by
\begin{eqnarray}
	P(E)=\frac{1}{C}\int_{I(E)}\sqrt{8\pi^2\gamma^2\rho_c
		\left[1-\frac{V(\phi)}{\rho_c}\right]}d\phi,\label{pro}
\end{eqnarray} 
in which $I(E)$ indicates the integral limits correspond to the range of $\phi$ that generates the desired slow-roll inflation \cite{Graef}, and the normalization $C$ is given by
\begin{eqnarray}
	C=\int^{\phi_{max}^0}_{\phi_{min}^0}\sqrt{8\pi^2\gamma^2\rho_c \left[1-\frac{V(\phi)}{\rho_c}\right]}d\phi,\label{N}
\end{eqnarray}
For the quadratic potential, the maximum and minimum values of $\phi$ are determined by the condition $V(\phi_{\max}^0)=\rho_c$, and $\phi_{max}^0 = -\phi_{min}^0 \simeq 0.906/m$. Then, Eq. (\ref{pro}) can be expressed as
\begin{eqnarray}
	P(E)=\frac{\left.\left(\frac{\phi}{2} \sqrt{1.-1.221 m^2 \phi^2}+\frac{0.452 \sin ^{-1}(1.105 m \phi)}{m}\right)\right|_{I(E)}}{ \left.\left(\frac{\phi}{2}  \sqrt{1.-1.221 m^2 \phi^2}+\frac{0.452 \sin ^{-1}(1.105 m \phi)}{m}\right)\right|^{\frac{0.906}{m}}_{-\frac{0.906}{m}}}\label{pro1}
\end{eqnarray}
The integral range $I(E)$ can depend on the value of the mass parameter. Therefore, to investigate the influence of the mass on the inflationary dynamics, we consider different values of $m$. As shown in \eqref{prob_quadratic}, different mass values lead to different ranges of $I(E)$ and probability.

The critical energy density $\rho_c$ is the main parameter that characterizes the LQC quantum geometry effects, and we keep it fixed in our numerical simulations. The values $m=1.26\times10^{-6}m_{Pl}$ and $M=2.5\times10^{-6}m_{Pl}$ provide the best-fit values from the Planck 2015 observations under the standard slow-roll inflation scenario. However, these values are obtained without considering the effects of the pre-inflationary dynamics in LQC \cite{Li084029}. Therefore, in this work, we treat the mass parameters as free parameters in our numerical analysis. The relationship between the mass parameter and the number of e-folds will be investigated in the next section.

\section{Background evolutionary pictures}\label{Sec3}

Once the potential and the mass are specified, Eqs.~(\ref{Fri}) and (\ref{kG}) can be solved numerically by specifying the initial values of $a, H, \phi$ and $\dot\phi$ at a specific point in time. A convenient choice is the bounce point, denoted $t_B$. At the bounce, $H_B=0$ and the energy density $\rho_B=\rho_c$, so the four initial values $(a_B,H_B,\phi_B,\dot{\phi}_B)$ reduce to just $(a_B,\phi_B)$ or $(a_B,\dot{\phi}_B)$. We choose $(a_B,\phi_B)$ in this work. Note that we treat the values of $m$ and $M$ as free parameters and explore their effect on the background dynamics.

To discuss the evolution of the background, we first introduce the following background quantities:
\begin{description}
	\item{(1)} The ratio between $V(\phi_B)$ and $\rho_c$,
	\begin{eqnarray}
		F_B=\frac{V(\phi_B)}{\rho_c}.\label{FB}
	\end{eqnarray}
	This quantity lies in the range $F_B\in[0,1]$, since $V(\phi_B)\in[0,\rho_c]$. Following \cite{Sloan}, we divide the initial conditions into three cases: the extreme KED case for $F_B<10^{-4}$, the KED case for $10^{-4}<F_B<0.1$, and the PED case for $F_B>0.1$.
	\item{(2)} The equation of state (EoS), defined by
	\begin{eqnarray}
		\omega=\frac{P}{\rho},\label{omega}
	\end{eqnarray}
	where $P=\dot{\phi}^2/2-V(\phi)$ and $\rho=\dot{\phi}^2/2+V(\phi)$ are the pressure and the energy density of the scalar field, respectively. This parameter lies in the range $\omega\in[-1,1]$, so that in the KED regime $w=1$, and in the PED regime $w=-1$.
	\item{(3)} The slow-roll parameter $\varepsilon_H$, defined in terms of the Hubble parameter and its derivative as
	\begin{eqnarray}
		\varepsilon_H=-\frac{\dot{H}}{H^2}.
	\end{eqnarray}
	During the slow-roll inflation phase, $|\varepsilon_H| \ll 1$.
	\item{(4)} The e-fold number $N$, defined as
	\begin{eqnarray}
		N\equiv \ln \left(\frac{a_{end}}{a_{i}}\right)\label{N},
	\end{eqnarray}
	where $a_{end}$ ($a_i$) is the scale factor at the end (beginning) of a given phase. That is, $a_{end}^{SI}$ ($a_i^{SI}$), $a_{end}^{DM}$ ($a_i^{DM}$), and $a_{end}^{SL}$ ($a_i^{SL}$) denote the values of the scale factor at the end (beginning) of the super-inflation, damping, and slow-roll inflation phases, respectively. Clearly, $a_{end}^{SI}=a_i^{DM}=a(t_{SI})$, $a_{end}^{DM}=a_i^{SL}=a(t_{SL})$, and $a_{end}^{SL}=a(t_{end})$, where $t_{SI}$ is the time at which $\rho=\rho_c/2$, $t_{SL}$ is the time at which the Universe begins to accelerate, defined as the first time after the bounce that $|\epsilon_H(t_{SL})|=1$, and $t_{end}$ is the time at which $|\varepsilon_H(t_{end})| = 1$for the second time.
	
\end{description}

\begin{table*}
	\small
	\caption{Quantities involved in background evolution. The field with quadratic potential, and $\dot{\phi}_B>0,\phi_B>0$. The time $t_{\rm{si}}$ is the end of super-inflation, and $t_{\rm{end}}$ is the time of end of slow-roll inflation. The symbol 
	``--" means the value is very huge. The values of $F_B$ start from the PED case and approach the KED one.}
	\label{Tab1}
	\begin{tabular*}{\textwidth}{@{\extracolsep{\fill}}cccccccc@{}}
		\hline
		$m$ & $F_B$ & $t_{SI}$ & $N_{SI}$ & $t_{TR}$ & $N_{TR}$ & $t_{END}$ & $N$ \\
		\hline 
		\multirow{6}{*}{$1.2 \times 10^{-7}$} 
		& $0.95$   &  --  & --  & --  & --  & --  & --   \\
		& $0.1$    & $0.208$ & $0.135$ & $0.556$  & $0.419$ & --  & $>10^3$ \\
		& $10^{-4}$ & $0.180$ & $0.116$ & $20.63$  & $1.651$ & --  & $>10^3$ \\
		& $10^{-8}$ & $0.180$ & $0.116$ & $2.067 \times 10^3$ & $3.185$ & -- & $>10^3$ \\
		& $10^{-12}$ & $0.180$ & $0.116$ & $1.570\times 10^5$ & $4.627$ & $5.025 \times 10^8$ & $630.660$ \\
		& $10^{-15}$ & $0.180$ & $0.116$ & $5.526 \times 10^5$ & $5.041$ & $1.404 \times 10^8$ & $57.81$ \\
		\hline 
		\multirow{6}{*}{$1.2 \times 10^{-6}$} 
		& $0.95$   &  --  & --  & --  & --  & --  & --   \\
		& $0.1$    & $0.208$ & $0.135$ & $0.556$  & $0.419$ & --  & $>10^3$ \\
		& $10^{-4}$ & $0.180$ & $0.116$ & $20.63$  & $1.650$ & --  & $>10^3$ \\
		& $10^{-8}$ & $0.180$ & $0.116$ & $2.018 \times 10^3$ & $3.178$ & -- & $>10^3$ \\
		& $10^{-12}$ & $0.180$ & $0.116$ & $5.280\times 10^4$ & $4.259$ & $1.465 \times 10^7$ & $61.950$ \\
		& $10^{-15}$ & $0.180$ & $0.116$ & $6.881 \times 10^5$ & $4.345$ & $1.118 \times 10^7$ & $38.957$ \\
		\hline 
		\multirow{6}{*}{$1.2 \times 10^{-5}$} 
		& $0.95$   &  --  & --  & --  & --  & --  & --   \\
		& $0.1$    & $0.208$ & $0.135$ & $0.556$  & $0.419$ & --  & $>10^3$ \\
		& $10^{-4}$ & $0.180$ & $0.116$ & $20.63$  & $1.650$ & --  & $>10^3$ \\
		& $10^{-8}$ & $0.180$ & $0.116$ & $1.695 \times 10^3$ & $3.118$ & $4.648 \times 10^6$ & $540.$ \\
		& $10^{-12}$ & $0.180$ & $0.116$ & $7.936 \times 10^3$ & $3.624$ & $9.640 \times 10^5$ & $29.987$ \\
		& $10^{-15}$ & $0.180$ & $0.116$ & $8.219 \times 10^3$ & $3.635$ & $9.318 \times 10^5$ & $28.031$ \\
		\hline 
		\multirow{6}{*}{$1.2 \times 10^{-3}$} 
		& $0.95$   &  --  & --  & --  & --  & --  & --   \\
		& $0.1$    & $0.208$ & $0.135$ & $0.556$  & $0.419$ & --  & $>10^3$ \\
		& $10^{-4}$ & $0.180$ & $0.116$ & $18.433$  & $1.610$ & $4.269 \times 10^4$  & $456.515$ \\
		& $10^{-8}$ & $0.180$ & $0.116$ & $1.216 \times 10^2$ & $2.226$ & $6.223 \times 10^3$ & $14.077$ \\
		& $10^{-12}$ & $0.180$ & $0.116$ & $1.283 \times 10^2$ & $2.243$ & $5.388 \times 10^3$ & $12.988$ \\
		& $10^{-15}$ & $0.180$ & $0.116$ & $1.283 \times 10^2$ & $2.243$ & $5.884 \times 10^3$ & $12.977$ \\
		\hline 
	\end{tabular*}
\end{table*}
\begin{table*}
	\small
	\caption{Quantities involved in background evolution. The field with quadratic potential, and $\dot{\phi}_B>0,\phi_B<0$. The meaning of parameters/symbles are the same as the Table \ref{Tab1}.}
	\label{Tab2}
	\begin{tabular*}{\textwidth}{@{\extracolsep{\fill}}cccccccc@{}}
		\hline
		$m$ & $F_B$ & $t_{SI}$ & $N_{SI}$ & $t_{TR}$ & $N_{TR}$ & $t_{END}$ & $N$ \\
		\hline 
		\multirow{6}{*}{$1.2 \times 10^{-7}$} 
		& $0.95$   &  --  & --  & --  & --  & --  & --   \\
		& $0.1$    & $0.208$ & $0.135$ & $0.556$  & $0.419$ & --  & $>10^3$ \\
		& $10^{-4}$ & $0.180$ & $0.116$ & $20.630$  & $1.651$ & --  & $>10^3$ \\
		& $10^{-8}$ & $0.180$ & $0.116$ & $2.067 \times 10^3$ & $3.186$ & -- & $>10^3$ \\
		& $10^{-12}$ & $0.180$ & $0.116$ & $3.076 \times 10^5$ & $4.859$ & $2.482 \times 10^8$ & $162.939$ \\
		& $10^{-15}$ & $0.180$ & $0.116$ & $6.575 \times 10^5$ & $5.098$ & $1.170 \times 10^8$ & $42.858$ \\
		\hline 
		\multirow{6}{*}{$1.2 \times 10^{-6}$} 
		& $0.95$   &  --  & --  & --  & --  & --  & --   \\
		& $0.1$    & $0.208$ & $0.135$ & $0.556$  & $0.419$ & --  & $>10^3$ \\
		& $10^{-4}$ & $0.180$ & $0.116$ & $20.632$  & $1.651$ & --  & $>10^3$ \\
		& $10^{-8}$ & $0.180$ & $0.116$ & $2.109 \times 10^3$ & $3.193$ & -- & $>10^3$ \\
		& $10^{-12}$ & $0.180$ & $0.116$ & $1.007 \times 10^5$ & $4.468$ & $7.555 \times 10^6$ & $21.290$ \\
		& $10^{-15}$ & $0.180$ & $0.116$ & $7.019 \times 10^4$ & $4.351$ & $1.094 \times 10^7$ & $37.664$ \\
		\hline 
		\multirow{6}{*}{$1.2 \times 10^{-5}$} 
		& $0.95$   &  --  & --  & --  & --  & --  & --   \\
		& $0.1$    & $0.208$ & $0.135$ & $0.556$  & $0.419$ & --  & $>10^3$ \\
		& $10^{-4}$ & $0.180$ & $0.116$ & $20.655$  & $1.651$ & --  & $>10^3$ \\
		& $10^{-8}$ & $0.180$ & $0.116$ & $2.668 \times 10^3$ & $3.275$ & $2.876 \times 10^6$ & $213.732$ \\
		& $10^{-12}$ & $0.180$ & $0.116$ & $8.542 \times 10^3$ & $3.647$ & $8.945 \times 10^5$ & $26.586$ \\
		& $10^{-15}$ & $0.180$ & $0.116$ & $8.238 \times 10^3$ & $3.636$ & $9.291 \times 10^5$ & $28.206$ \\
		\hline 
		\multirow{6}{*}{$1.2 \times 10^{-3}$} 
		& $0.95$   &  --  & --  & --  & --  & --  & --   \\
		& $0.1$    & $0.208$ & $0.135$ & $0.556$  & $0.420$ & --  & $>10^3$ \\
		& $10^{-4}$ & $0.180$ & $0.116$ & $23.568$  & $1.644$ & $3.271 \times 10^4$  & $271.791$ \\
		& $10^{-8}$ & $0.180$ & $0.116$ & $1.358 \times 10^2$ & $2.261$ & $5.573 \times 10^3$ & $11.954$ \\
		& $10^{-12}$ & $0.180$ & $0.116$ & $1.284 \times 10^2$ & $2.243$ & $5.881 \times 10^3$ & $12.966$ \\
		& $10^{-15}$ & $0.180$ & $0.116$ & $1.284 \times 10^2$ & $2.243$ & $5.844 \times 10^3$ & $12.977$ \\
		\hline 
	\end{tabular*}
\end{table*}

\begin{table*}
	\small
	\caption{Quantities involved in background evolution. The field with Starobinsky potential, and $\dot{\phi}_B>0$. The meaning of parameters are the same as the Table \ref{Tab1}. The values of $F_B$ start from the PED case and approach the KED one.}
	\label{Tab3}
	\begin{tabular*}{\textwidth}{@{\extracolsep{\fill}}cccccccc@{}}
		\hline
		$M$ & $F_B$ & $t_{SI}$ & $N_{SI}$ & $t_{TR}$ & $N_{TR}$ & $t_{END}$ & $N$ \\
		\hline 
		\multirow{6}{*}{$2.4 \times 10^{-7}$} 
		& $0.95$   & $0.404$ & $0.202$ & \multicolumn{2}{c}{No transition} & \multicolumn{2}{c}{No inflation} \\
		& $0.1$    & $0.193$ & $0.123$ & \multicolumn{2}{c}{No transition} & \multicolumn{2}{c}{No inflation} \\
		& $10^{-3}$ & $0.180$ & $0.116$ & \multicolumn{2}{c}{No transition} & \multicolumn{2}{c}{No inflation} \\
		& $10^{-5}$ & $0.180$ & $0.116$ & $4.580 \times 10^6$ & $5.370$ & $2.292 \times 10^7$ & $7.299$ \\
		& $10^{-8}$ & $0.180$ & $0.116$ & $3.221 \times 10^6$ & $5.634$ & $6.434 \times 10^8$ & $79.269$ \\
		& $10^{-9}$ & $0.180$ & $0.116$ & $3.195 \times 10^6$ & $5.633$ & $2.011 \times 10^9$ & $242.483$ \\
		\hline 
		\multirow{6}{*}{$2.4 \times 10^{-6}$} 
		& $0.95$   & $0.404$ & $0.202$ & \multicolumn{2}{c}{No transition} & \multicolumn{2}{c}{No inflation} \\
		& $0.1$    & $0.193$ & $0.123$ & \multicolumn{2}{c}{No transition} & \multicolumn{2}{c}{No inflation} \\
		& $10^{-3}$ & $0.180$ & $0.116$ & \multicolumn{2}{c}{No transition} & \multicolumn{2}{c}{No inflation} \\
		& $10^{-5}$ & $0.180$ & $0.116$ & $3.777 \times 10^5$ & $4.898$ & $4.811 \times 10^6$ & $8.936$ \\
		& $10^{-8}$ & $0.180$ & $0.116$ & $3.201 \times 10^5$ & $4.866$ & $1.367 \times 10^8$ & $164.813$ \\
		& $10^{-9}$ & $0.180$ & $0.116$ & $3.189 \times 10^5$ & $4.866$ & $4.240 \times 10^8$ & $508.624$ \\
		\hline 
		\multirow{6}{*}{$2.4 \times 10^{-5}$} 
		& $0.95$   & $0.404$ & $0.202$ & \multicolumn{2}{c}{No transition} & \multicolumn{2}{c}{No inflation} \\
		& $0.1$    & $0.193$ & $0.123$ & \multicolumn{2}{c}{No transition} & \multicolumn{2}{c}{No inflation} \\
		& $10^{-3}$ & $0.180$ & $0.116$ & \multicolumn{2}{c}{No transition} & \multicolumn{2}{c}{No inflation} \\
		& $10^{-5}$ & $0.180$ & $0.116$ & $3.446 \times 10^4$ & $4.111$ & $9.908 \times 10^5$  & $13.794$ \\
		& $10^{-8}$ & $0.180$ & $0.116$ & $3.192 \times 10^4$ & $4.098$ & $2.773 \times 10^7$ & $332.$ \\
		& $10^{-9}$ & $0.180$ & $0.116$ & $3.187 \times 10^4$ & $4.098$ & $7.724 \times 10^7$ & $925.531$ \\
		\hline 
		\multirow{6}{*}{$2.4 \times 10^{-3}$} 
		& $0.95$   & $0.404$ & $0.202$ & \multicolumn{2}{c}{No transition} & \multicolumn{2}{c}{No inflation} \\
		& $0.1$    & $0.193$ & $0.123$ & \multicolumn{2}{c}{No transition} & \multicolumn{2}{c}{No inflation} \\
		& $10^{-3}$ & $0.180$ & $0.116$ & $3.790 \times 10^2$  & $2.596$ & $4.726 \times 10^3$  & $6.541$ \\
		& $10^{-5}$ & $0.180$ & $0.116$ & $3.249 \times 10^2$ & $2.566$ & $3.674 \times 10^4$  & $43.496$ \\
		& $10^{-8}$ & $0.180$ & $0.116$ & $3.196 \times 10^2$ & $2.563$ & $1.836 \times 10^5$ & $218.540$ \\
		& $10^{-9}$ & $0.180$ & $0.116$ & $3.195 \times 10^2$ & $2.563$ & $2.020 \times 10^5$ & $240.499$ \\
		\hline 
	\end{tabular*}
\end{table*}

The numerical results are presented in Tables \ref{Tab1}, \ref{Tab2}, and \ref{Tab3}. It is easy to see that the main properties remain the same for different choices of the mass, a result consistent with those found in modified LQC \cite{Li066016}. We now analyze the numerical results in detail.

As mentioned in Sec.~\ref{Sec2}, at the beginning of the expanding phase, the universe enters a super-inflationary phase right after the quantum bounce. In the KED case, the duration of the super-inflation phase does not depend on the initial condition or on the type of potential. However, in the PED case, the duration depends on the initial condition and the potential. For instance, at $F_B=0.95$, the super-inflation phase lasts for more than $10^5\,t_{Pl}$ for the quadratic potential, while for the Starobinsky potential it lasts only for a short period. The numerical results for both cases are given in Tables \ref{Tab1}, \ref{Tab2}, and \ref{Tab3}. The damping phase is the stage between the super-inflation phase and the slow-roll inflation phase. The duration of this phase also depends on the type of potential and on the initial condition. In the quadratic case, this phase does not exist for $F_B=0.95$, since $\omega$ remains smaller than $-1/3$ throughout the super-inflation stage.

The slow-roll inflation phase begins when $|\epsilon_H|=1$, for the first time after the bounce, and ends when $|\epsilon_H|=1$; during this phase, the Hubble slow-roll parameter satisfies $\varepsilon_H\ll1$ indicating a positive acceleration $\ddot{a} > 0$. As Tables \ref{Tab1}, \ref{Tab2}, and \ref{Tab3} show, the duration of this phase, and hence the resulting e-fold number, depends on the type of potential, mass, and the initial condition. We consider four different values of the mass for each potential, $m\,(M)=10^{-7} m_{Pl}$, $10^{-5} m_{Pl}$, $10^{-3} m_{Pl}$. In the quadratic case, as in \cite{Sloan2}, it is easy to obtain a sufficient number of e-folds when the initial condition is of the PED type\footnote{In fact, for the case of extreme PED case, number of e-folds during the inflationary phase easily exceeds $10^3$.}. 
In the Starobinsky case, as in \cite{Bonga1}, the KED case is more suitable for obtaining enough e-folds, no inflationary phase is generated for the PED case. For the quadratic potential, a smaller mass makes it easier to reach a sufficient number of e-folds, as shown in Tables \ref{Tab1}, \ref{Tab2}; however, for the Starobinsky potential, larger mass results in higher e-folds, shown in Table \ref{Tab3}.

For extreme KED initial conditions, both potentials generically go through all three phases, the super-inflation, the damping, and the slow-roll inflation phase, so that the total amount of expansion depends on the values of $m$ and $F_B$. On the other hand, for the most extreme PED initial conditions, the two potentials behave very differently. For the Starobinsky potential, neither the damping phase nor the slow-roll inflation phase takes place. However, for the quadratic potential, both phases do occur, and the resulting e-fold number easily exceeds $10^3$. In between these two limits, for the quadratic potential with $F_B \geq 0.5$, the slow-roll inflation phase begins before the super-inflation phase ends, so that the damping phase does not exist and the slow-roll phase itself unfolds while quantum effects are still dominant. The case $F_B=0.95$ can be taken as an example such that the entire slow-roll phase takes place in the quantum-dominated regime, the super-inflation phase ends only after $t>10^5\,t_{Pl}$, and the resulting e-fold number is extremely large.

We next investigate the probability of realizing slow-roll inflation for the quadratic potential. Using Eq. (\ref{pro1}), we obtain
\begin{equation}\label{prob_quadratic}
	P(E)  \left\{
	\begin{array}{cl}
		\simeq 1.000000 \quad &\rm{for}\quad m = 10^{-7}m_{Pl}, \\
		=0.999994 \quad  &\rm{for}\quad m = 1.360\times10^{-6}m_{Pl}, \\
		=0.999953 \quad &\rm{for}\quad m = 10^{-5}m_{Pl},\\
		=0.997295 \quad  &\rm{for}\quad m = 10^{-3}m_{Pl}.
	\end{array}
	\right.
\end{equation}
These results indicate that the probability of obtaining a slow-roll inflationary phase increases as the mass parameter decreases.

In this section, we discussed the relationship between the mass and the background evolution. We found that the evolutionary picture of the background depends on the type of the potentials, the initial condition and the value of the mass. In the next section, as an example, we will investigate the primordial power spectra. In the rest of the work, we focus on the branch with $\dot{\phi}_B>0$. The case with $\dot{\phi}_B<0$ can be analyzed in a similar way and leads to the same qualitative behavior. In the numerical analysis, the mass parameter is treated as a free parameter of the model. 

\section{Primordial power spectra}\label{Sec4}

The background solutions were obtained in Sec.~\ref{Sec3} for the two potentials considered in this work. They show how the volume, the scalar field, and the other relevant quantities behave around the bounce, and how the slow-roll inflation is generated naturally. To connect the theory with the observations of the Cosmic Microwave Background (CMB), one has to study the linear perturbations that propagate on top of this quantized background. Due to the ambiguities in the loop quantization of the perturbed sector, several approaches to the cosmological perturbations have been proposed in the LQC framework. The dressed metric approach and the hybrid approach have been widely used in the literature. More recently, an alternative mass function approach has also been introduced, based on a different polymerization of the classical mass term \cite{LiSingh2024mass}. We first briefly introduce these methods and provide the main features. More detail information can be found in \cite{LiSingh2022,Lietal_constraining,LiSingh2024mass,MohammadiLiZhu2025}.

\subsection{Three approaches to cosmological perturbations in LQC}\label{Sec4a}

The linear perturbations in all three methods follow the modified Mukhanov-Sasaki equation in which only the effective mass functions differs from one approach to another. These differences originate from the quantum ambiguity in treating the perturbed sector. The Mukhanov-Sasaki equation is given by \footnote{following the notation given in \cite{LiSingh2024mass}}
\begin{equation}\label{eq:MS}
	\nu_k''+\left(k^2+s\right)\nu_k=0,
\end{equation}
in which $\nu_k$ is the rescaled variable related to the comoving curvature perturbations by $\nu_k=z_s\mathcal{R}_k$ where $z_s\equiv a\dot\phi/H$. The prime indicates derivative with respect to the conformal time $\eta$, defined by $a\,d\eta=dt$. The effective mass function is denoted by $s$, and its explicit form depends on which of the three approaches is considered. We solve Eq.~(\ref{eq:MS}) using the background solution obtained in Sec.~\ref{Sec3}, and we set the Bunch-Davies vacuum state as the initial condition for the mode functions. The resulting power spectrum is estimated as 
\begin{equation}\label{eq:Ps}
	\mathcal{P}_s = \frac{k^3}{2 \pi^2} \Big| \frac{\nu_k}{z_s} \Big|^2
\end{equation}
This primordial power spectrum is subsequently fed into the CAMB code, and the resulting angular power spectrum is compared with the Planck data.

In the following lines, we are going to introduce the effective mass functions for dressed metric, hybrid and the alternative mass function approaches. 

\textit{Dressed metric approach.} The dressed metric approach is based on the Hamiltonian formulation of cosmological perturbation theory introduced by Langlois \cite{Langlois1994}. In this method, the Hamiltonian constraint is first expanded to second order in the perturbation variables. The zeroth-order part governs the background and is loop quantized, in the same way as in Sec.~\ref{Sec2}. The second-order part then describes a Fock-quantized field propagating on this quantum-corrected geometry \cite{AgulloDM1,AgulloDM2,AgulloDM3}.

The classical mass term is given by
\begin{eqnarray}\label{DMmass}
	s & = & U^2-\frac{a''}{a}, \\
	U^2 & = & \frac{24\pi G\,p_\phi^2}{a^4}-\frac{18\,p_\phi^4}{a^6\pi_a^2}-\frac{12\,a\,p_\phi\,V_{,\phi}}{\pi_a}+a^2V_{,\phi\phi}, \nonumber 
\end{eqnarray}
where $\pi_a$ is the momentum conjugate to the scale factor, and $V_{,\phi}$ and $V_{,\phi\phi}$ are the first and second derivatives of the inflaton potential. Since $\pi_a$ is classically related to the connection-like variable $b$ through $\pi_a=-6a^2b/(8\pi G\gamma)$, the terms $1/\pi_a$ and $1/\pi_a^2$ in Eq.~(\ref{DMmass}) must also be polymerized, following the same rule $b\to\sin(\lambda b)/\lambda$ used for the background \cite{MaruganHA3,MaruganHA2,LiWangSingh2020DM}. In LQC, $1/\pi_a$ and $1/\pi_a^2$ are polymerized as
\begin{equation}\label{DMpoly}
	\frac{1}{\pi_a}\to-\frac{4\pi G\gamma\lambda\cos(\lambda b)}{3a^2\sin(\lambda b)}, \qquad
	\frac{1}{\pi_a^2}\to\frac{16\pi^2G^2\gamma^2\lambda^2}{9a^4\sin^2(\lambda b)}.
\end{equation}
Substituting the polymerization \eqref{DMpoly} in \eqref{DMmass}, one can find the effective mass function for the dressed metric approach.

\textit{Hybrid approach.} The hybrid approach follows the same hybrid quantization scheme as the dressed metric approach, with the background loop quantized and the perturbations Fock quantized \cite{MaruganHA1,MaruganHA2,MaruganHA3,MaruganHA4}. The classical mass function is written in a form that is classically equivalent to the one used in the dressed metric approach, but polymerization results in a different effective mass function \cite{LiOlmedoSingWang2020Hyb,LiSingh2022}. This is the origin of the difference between the dressed metric and the hybrid predictions for the power spectrum. The effective mass function in the hybrid approach is given by
\begin{eqnarray}\label{Hybridmass}
	s & = & \frac{4\pi G\,p_\phi^2}{3v^{4/3}}\left(19-\frac{24\pi G\gamma^2p_\phi^2}{\Omega^2}\right) \\
	& & \qquad +v^{2/3}\left(V_{,\phi\phi}+\frac{16\pi G\gamma\,p_\phi\Lambda}{\Omega^2}V_{,\phi}-\frac{16\pi G}{3}V\right), \nonumber
\end{eqnarray}
For the standard LQC, the variables $\Omega$ and $\Lambda$, equal to each other classically, are polymerized as
\begin{equation}
	\Omega = v\,\frac{\sin(\lambda b)}{\lambda}, \qquad \Lambda = v\,\frac{\sin(2\lambda b)}{2\lambda}.
	\label{Hybridpoly}
\end{equation}

\textit{Alternative mass function.} With the third approach, the story starts from the comoving gauge of the Lagrangian formulation of perturbation theory, where the classical mass function takes the compact form
\begin{equation}
	m_{\rm CG}^2 = -\frac{z_s''}{z_s}.
	\label{CGmass}
\end{equation}
Instead of polymerizing $\pi_a$ or $b$ directly, one can polymerize the inverse Hubble rate that appears inside $z_s=a\dot\phi/H$, through the substitution
\begin{equation}\label{CGpoly}
	z_s \;\to\; \frac{a\dot\phi\sqrt{\rho_c-\rho}}{H\,g(\rho)},
\end{equation}
where $g(\rho)$ is a free function which generally must satisfied two constraints as $g(\rho)|_{\rho\to\rho_c}={\rm const}$ and $g(\rho)|_{\rho\ll\rho_c}\to1/\sqrt{\rho_c}$ \cite{LiSingh2024mass}. This produces an effective mass of the form $m_{\rm eff}^2=\Omega_g^2-a''/a$, with
\begin{eqnarray}\label{OmegaG}
	\Omega_g^2 & = & a^2\bigg[V_{,\phi\phi}+\left(48\pi G+\delta_a\right)V+\left(\frac{6H\dot\phi}{\rho}+\delta_b\right)V_{,\phi} \nonumber\\
	& & \qquad +\left(\delta_c-\frac{48\pi G}{\rho}\right)V^2+\delta_d\,\rho^2\bigg],
\end{eqnarray}
containing four extra correction terms relative to Eq.~(\ref{DMmass}), all built out of $g(\rho)$ and its derivatives with respect to $\rho$,
\begin{eqnarray}
	\delta_a & = & -\frac{24\pi G\rho}{\rho_c\,g}\Big(g-2(\rho_c-3\rho)g_{,\rho}+8\rho(\rho-\rho_c)g_{,\rho\rho}\Big), \label{deltaa}\\
	\delta_b & = & -12H\dot\phi\,\frac{g_{,\rho}}{g}, \label{deltab}\\
	\delta_c & = & \frac{24\pi G}{\rho_c\,g}\Big(g+2\rho_c\,g_{,\rho}+4\rho(\rho-\rho_c)g_{,\rho\rho}\Big), \label{deltac}\\
	\delta_d & = & -\frac{48\pi G}{\rho_c\,g}\Big((2\rho_c-3\rho)g_{,\rho}+2\rho(\rho_c-\rho)g_{,\rho\rho}\Big). \label{deltad}
\end{eqnarray}
Because the map from full LQG to the cosmological sector is not yet settled, $g(\rho)$ is left as a phenomenological input. Following Ref.~\cite{LiSingh2024mass} and truncating it to linear order in $\rho/\rho_c$,
\begin{equation}\label{gansatz}
	g(\xi;\rho) = \frac{1}{\sqrt{\rho_c}}\left(1+\xi\,\frac{\rho}{\rho_c}\right),
\end{equation}
introduces a single dimensionless parameter $\xi$, which determines how the correction terms in the effective mass function depend on the energy density near the bounce. No choice of $\xi$ reduces this mass function to the one used in the dressed metric or the hybrid approach \cite{LiSingh2024mass}. 

All three effective mass functions are equivalent in the classical limit, so the dressed metric, hybrid, and alternative mass function schemes only disagree with one another in the neighborhood of the bounce. Since modes that are affected by this regime are precisely those that re-enter the horizon at the largest angular scales today, the observational imprint of these three quantization choices should be sought at low multipoles of the CMB angular power spectrum, which is what we compute and compare in Sec.~\ref{Sec4b}.

\subsection{Primordial tensor perturbations in LQC}\label{Sec4a}
We calculate the tensor power spectrum using the dressed metric approach, following the treatment in \cite{zhu-preinflation2}. The tensor perturbation mode function $Q_k$ satisfies the equation
\begin{equation}\label{eq:tensor_perturbation}
	\ddot{Q}_k+3H\dot{Q}_k+\frac{k^2}{a^2}Q_k=0 ,
\end{equation}
where the quantum gravity effects are included through the LQC-corrected background evolution. After solving Eq.~(\ref{eq:tensor_perturbation}), the tensor power spectrum is evaluated as
\begin{equation}\label{eq:Pt}
	\mathcal{P}_{T}(k)=\frac{k^3}{\pi}\left|Q_k\right|^2 .
\end{equation}
After solving Eq.~(\ref{eq:tensor_perturbation}) for a range of $k$ modes, we obtain the tensor power spectrum and together with the estimated scalar power spectrum, one can obtain the tensor-to-scalar ratio at the pivot scale $k_\star$.

\subsection{Numerical results}\label{Sec4b}

To get the power spectrum, we numerically solve the modified Mukhanov-Sasaki equation, Eq.\eqref{eq:MS}, using the background solution explained in Sec.~\ref{Sec3} for the corresponding potential and mass. The Bunch-Davies initial condition is set at the contracting phase, sufficiently before the bounce where all the modes of interest are inside the horizon. Then, the mode equation is evolved numerically up to the inflationary phase when the modes exit the horizon and remain deep outside the horizon and freeze. The power spectrum is estimated through Eq.\eqref{eq:Ps}. We first study how the mass of the inflaton affects the resulting power spectrum, before focusing on the particular mass value that reproduces the observed amplitude of the spectrum.

\begin{figure*}[t]
	\centering
	\subfigure{\includegraphics[width=0.95\linewidth]{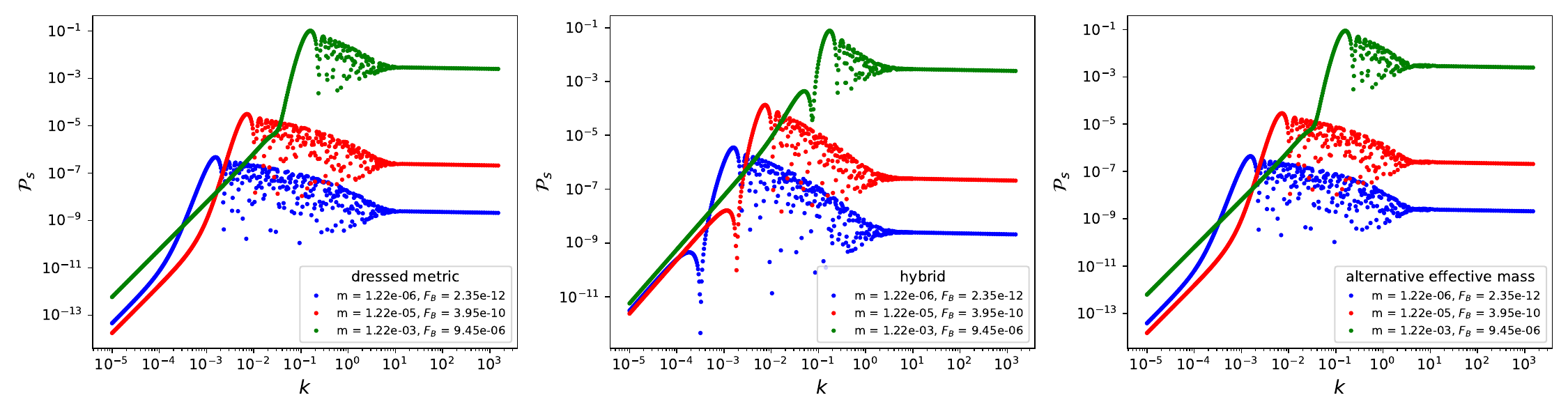}}
	\subfigure{\includegraphics[width=0.95\linewidth]{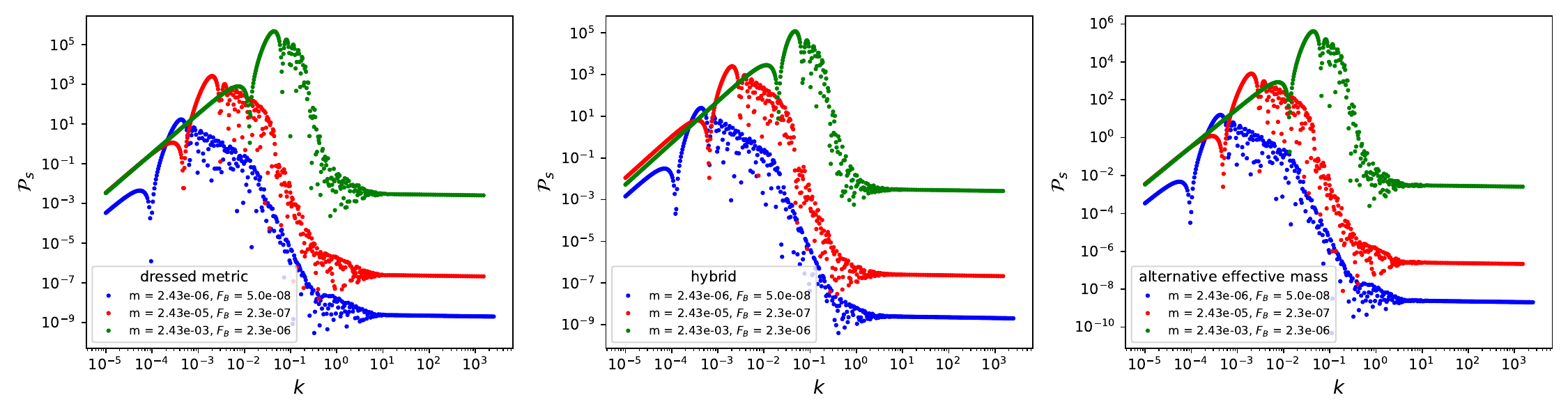}}
	\caption{The figure shows the resulting power spectra estimated by numerically solving the Mukhanov-Sasaki equation for quadratic potential (top row) and Starobinsky potential (bottom row). It shows the resulting power spectrum for different inflaton mass indicating the effect of mass on the power spectrum. }
	\label{fig:Psk_mF}
\end{figure*}
To get the numerical solution, the inflaton mass $m$ and $F_B$ parameter for both potentials are fixed such that the total number of e-folds from the bounce to the end of inflation is around $N \simeq 75$. As it will be discussed later, this chose guarantees that the pivot mode $k_\star/a_0=0.05\,{\rm Mpc}^{-1}$ exits the horizon roughly $55$ e-folds before the end of inflation, as required by observations \cite{liddle_2003,Baumann_Tasi}. We investigate how the mass of the inflaton affects the power spectrum. For this purpose, for each potential we calculate the power spectrum for three different mass. The results are presented in Fig.\ref{fig:Psk_mF}, with top row is the power spectrum for the quadratic potential, and the bottom row shows the resulting power spectrum for the Starobinsky potential. The overall shape of the power spectrum, the suppressed infrared regime, the oscillatory intermediate regime, and the scale-invariant regime, remains the same for every mass and every approach, as expected, since the mass does not change the qualitative dynamics near the bounce. What actually does change is the amplitude of the scale invariant regime, which increases with mass of the inflaton, and the intermediate regime which gets narrower. Among the mass values considered above, one can find that the mass with the order of $m \sim 10^{-6}$ can lead to a scale-invariant amplitude consistent with the observed value of the power spectrum. We now focus on this mass value and discuss the resulting power spectrum in more detail.

	\begin{figure*}[!h]
		\centering
		\subfigure{\includegraphics[width=0.45\linewidth]{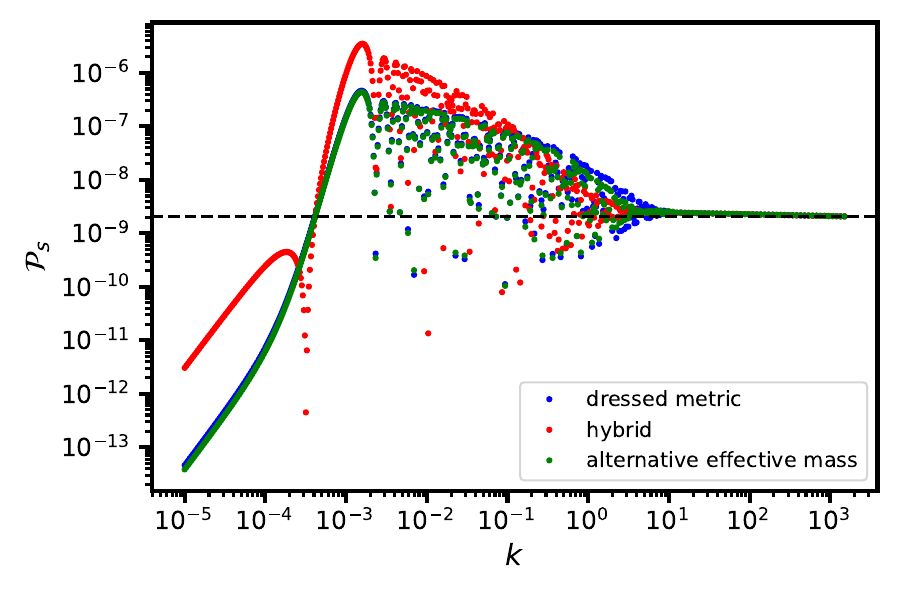}}
		\subfigure{\includegraphics[width=0.45\linewidth]{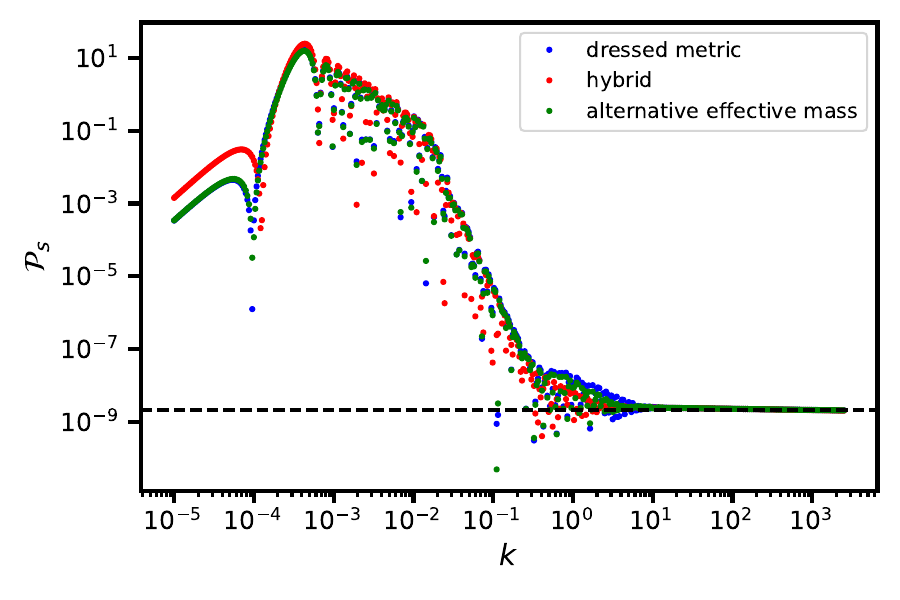}}
		\caption{The figure shows the resulting power spectra estimated by numerically solving the Mukhanov-Sasaki equation using three approaches: a) dressed metric (blue dots), b) hybrid approach (red dots), and c) alternative mass function (green dots).}
		\label{fig:Psk}
	\end{figure*}
To fix the exact value of the mass for each potential, we impose the additional requirement that the resulting scale-invariant amplitude match the observed value $A_s=2.0989\times10^{-9}$ of the power spectrum. In this regard, for the quadratic potential we have $m=1.22\times10^{-6}$ and $F_B = 2.3 \times 10^{-12}$, and for the Starobinsky potential the parameters are set as $m=2.43\times10^{-6}$ and $F_B = 5 \times 10^{-8}$. With these choices of the parameters, for both potentials, we obtain a total number of e-folds $N\simeq75$ from the bounce point to the end of inflation, which, as we show later, places the pivot mode at the horizon-crossing point of interest. In addition, these choices correspond to a kinetic-dominated bounce, from which the inflationary phase arises naturally. The resulting power spectra, using three approaches, are presented in Fig.\ref{fig:Psk}, with the quadratic potential in the left panel and the Starobinsky potential in the right panel. It is realized from the figure that, the resulting power spectra for all three approach and for both potentials, has quantitatively the same pattern so that the power spectrum is suppressed at small $k$ (the infrared regime), it is amplified and oscillates over an intermediate range of $k$, and it then shows a nearly scale-invariant behavior at large $k$, consistent with the observed spectrum on CMB scales\footnote{As discussed in \cite{LiSingh2024mass}, there is extra feature for the alternative mass approach in which there is a wavepacket shape like located just before the scale invariant regime. }. The three approaches differ mainly in how quickly the enhances oscillatory part converges into the scale invariant regime. The hybrid approach reaches the scale-invariant regime at the smallest value of $k$, followed by the alternative mass function approach. The dressed metric approach requires the largest $k$ to converge. These general behaivor are similar for both potentials. Additionally, it can be observed that the amplitude of the power spectra for the quadratic potential in the infrared and intermediate regimes are higher for the Starobinsky potential. To estimate the pivot scale, we followed the same procedure used in Refs.~\cite{Lietal_constraining,MohammadiLiZhu2025} and fix the pivot scale $k_\star$ in each case by matching the amplitude of the power spectrum in the scale-invariant regime to the observed value $\mathcal{P}_\mathcal{R}(k_\star)=A_s=2.0989\times10^{-9}$. 
For the quadratic potential, the pivot scale is obtained as $k_\star=1084.67$, $1037.31$, and $1036.59$ for the dressed metric, the hybrid, and the alternative mass function approaches, respectively, and for the Starobinsky potential we obtain $k_\star=1009.09$, $873.15$, and $1064.52$. 

\begin{figure}
	\centering
	\includegraphics[width=0.5\linewidth]{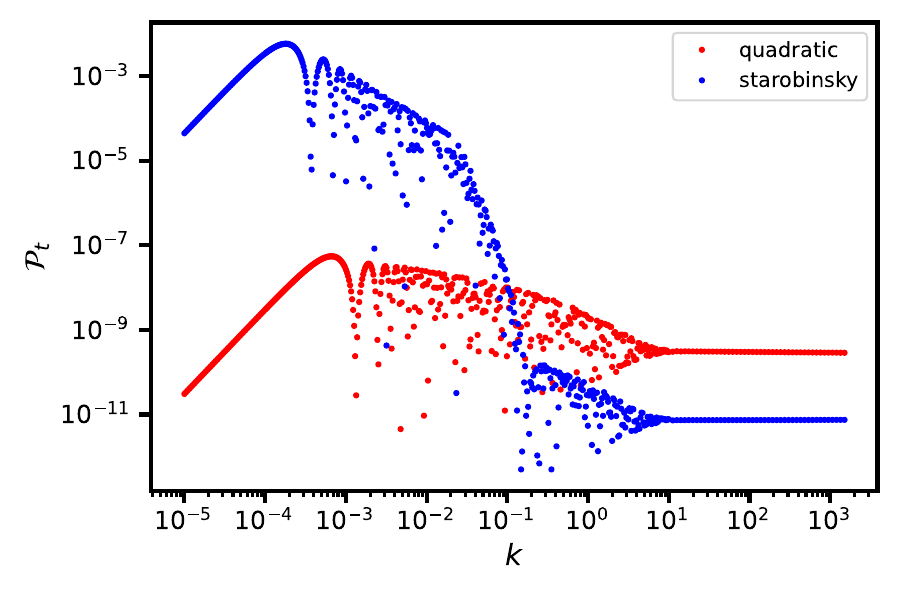}
	\caption{The figure presents the primordial tensor power spectrum for quadratic potential (red dots) and Starobinsky potential (blue dots) versus the $k$ modes.}
	\label{fig:Ptk}
\end{figure}
Numerically solving the tensor perturbation equation \eqref{eq:tensor_perturbation}, following the same method as for the scalar perturbation, one can obtain the tensor power spectrum. Fig.\ref{fig:Ptk} represents the resulting power spectrum for both quadratic and Starobinsky potential. As it can be observed, the power spectrum becomes almost scale invariant in our interest $k$ modes. For the quadratic potential, the spectrum at the scale invariant regime is around $5.9 \times 10^{-12}$ and it is around $1.45 \times 10^{-12}$ for the Starobinsky potential. Having the scalar and tensor power spectra, one can estimate the values of the scalar spectral index and the tensor-to-scalar ratio. Table.\ref{tab:rns} shows the resulting scalar spectral index obtained from the three approaches for both potentials. Additionally, the tensor-to-scalar ratio is obtained as $r = 0.13$ and $3.54 \times 10^{-3}$ for quadratic and Starobinsky potential respectively. The result for both the scalar spectral index and the tensor-to-scalar ratio shows a good agreement with the data for the Starobinsky potential.
\begin{table}[h]
	\centering
	\caption{The resulting scalar spectral index, $n_s - 1 = d\ln(\mathcal{P}_s)/dln(k)$, for all three approaches and both potentials. The abbreviations ``DM'', ``Hy'', and  ``AMF'' stand for dressed metric, hybrid and alternative mass function approaches. }
	\label{tab:rns}
	\begin{tabular}{c|ccc}
		\hline
		\text{Potential} & DM & Hy & AMF \\
		\hline
		quadratic & $0.9673$ & $0.9661$ & $0.9656$ \\
		\hline
		Starobinsky & $0.9670$ & $0.9663$ & $0.9689$ \\
		\hline
	\end{tabular}
\end{table}

Table~\ref{tab:kpivot} shows the effect of inflaton mass on the pivot scale $k_\star$. It presents the resulting $k_\star$ for three values of mass all in the same order, for the three perturbation approaches and for both potentials. In all cases, $k_\star$ increases with increasing mass, consistent with what is observed in Fig.~\ref{fig:Psk_mF}. This change is actually quite sharp so that for the quadratic potential, changing the mass by about $9\%$, from $m=1.12\times10^{-6}$ to $m=1.32\times10^{-6}$, is enough to more than double the value of $k_\star$ in all three approaches. This effect even gets stronger for the Starobinsky potential where a change of only about $2\%$ in the mass, from $m=2.38\times10^{-6}$ to $m=2.48\times10^{-6}$, the pivot scale $k_\star$ changes by one order of magnitude. This shows that the pivot scale is highly sensitive to the precise value of the inflaton mass. Comparing the three approaches at fixed mass, the dressed metric approach generally gives the largest pivot scale and the hybrid approach the smallest, in agreement with the pattern already found in Fig.~\ref{fig:Psk}, although the alternative mass function approach is slightly larger than the dressed metric approach for the Starobinsky potential at $m=2.43\times10^{-6}$.
\begin{table}[]
	\centering
	\caption{The table shows the effect of mass on the predicted pivot scale for the LQC model. The parameter $F_B$ is chosen as $2.35 \times 10^{-12}$ for the quadratic potential and $5 \times 10^{-8}$ for the Starobinsky potential. It is seen that the pivot scale increases with increasing the mass. The superscripts ``DM'', ``Hy'', and  ``AMF'' stand for dressed metric, hybrid and alternative mass function approaches. }
	\label{tab:kpivot}
	\begin{tabular}{c|cccc}
		\text{Potential}   & $m$ & $k_\ast^{DM}$ & $k_\ast^{Hy}$ & $k_\ast^{AMF}$ \\
		\hline
		\multirow{3}{*}{quadratic} & $1.12 \times 10^{-6}$ & $679.62$ & $661.45$ & $670.20$ \\
		& $1.22 \times 10^{-6}$ & $1084.67$ & $1037.31$ & $1036.59$ \\
		& $1.32 \times 10^{-6}$ & $1742.12$ & $1669.56$ & $1654.91$ \\
		\hline
		\multirow{3}{*}{Starobinsky} & $2.38 \times 10^{-6}$ & $169.48$ & $151.83$ & $158.62$ \\
		& $2.43 \times 10^{-6}$ & $1009.09$ & $873.15$ & $1064.52$ \\
		& $2.48 \times 10^{-6}$ & $2286.84$ & $1749.29$ & $1980.56$ \\
		\hline
	\end{tabular}
\end{table}

\begin{figure*}[!h]
		\centering
		\subfigure{\includegraphics[width=0.45\linewidth]{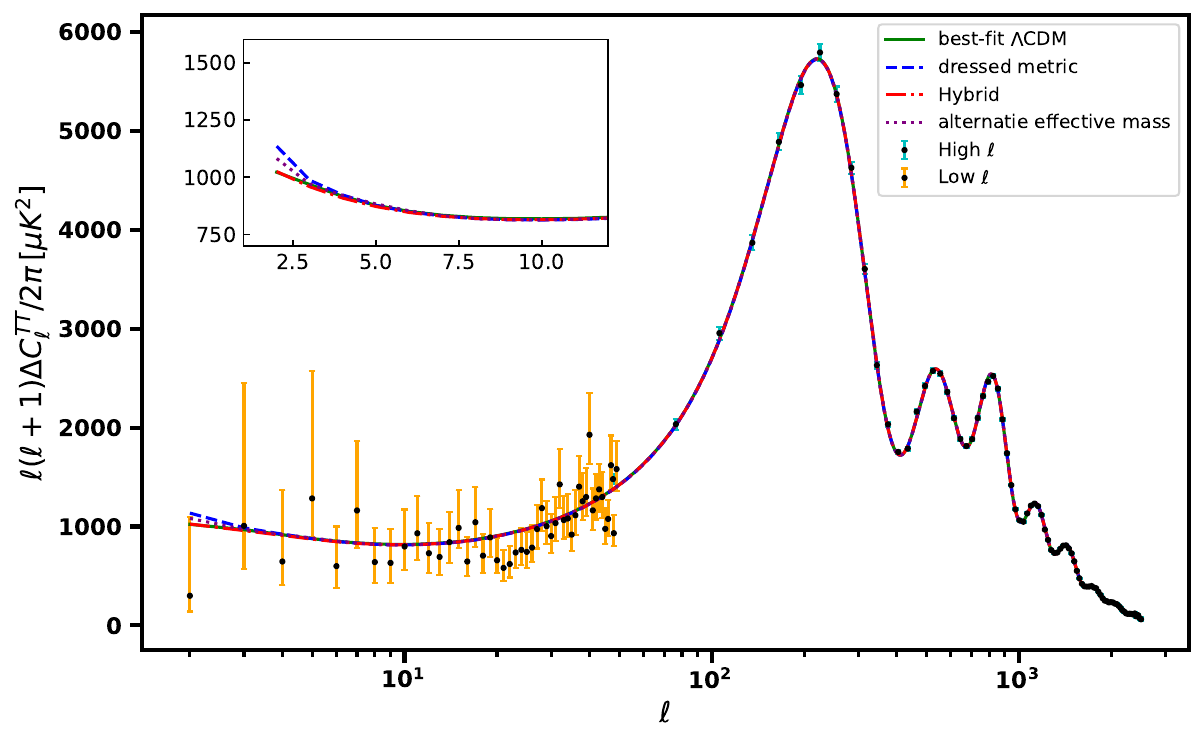}}
		\subfigure{\includegraphics[width=0.45\linewidth]{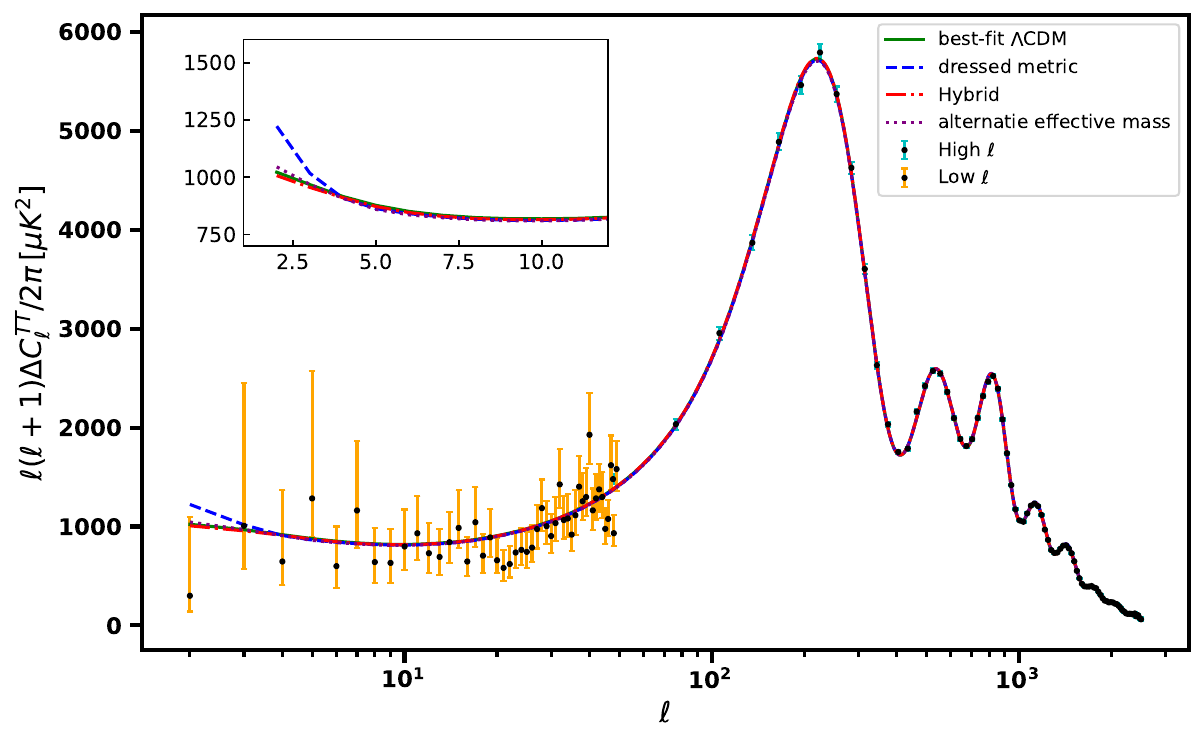}}
		\caption{The resulting angular power spectrum for all three approach for quadratic potential (left panel) and Starobinsky potential (right panel). }
		\label{fig:angular}
	\end{figure*}

Once the pivot scale is determined for each approach and potential, as estimated above, the resulting primordial power spectrum is fed as input to the CAMB code \cite{CAMB}, to produce the temperature angular power spectrum $C_l^{TT}$. 
The resulting angular power spectrum is compared with the Planck 2018 temperature data and with the best-fit $\Lambda$CDM curve obtained from the Planck likelihood \cite{Planck2020,Planck2018}, presented in Fig.\ref{fig:angular}, with interest in the low-multipole region where any imprint of the quantum bounce is expected to be most visible. It is observed from Fig.\ref{fig:angular} that, the resulting angular power spectrum for all three approaches, the dressed metric, the hybrid, and the alternative effective mass approaches, and for both potentials are a perfect match with the data. This agreement at high multipoles follows directly from the way $k_\star$ is fixed. Matching the primordial amplitude to the observed value inside the scale-invariant regime, as performed above, guarantees that the angular power spectrum agrees with the $\Lambda$CDM prediction at these scales \cite{LiWangSingh2020DM}. The differences between approaches, and also between the potential, only become visible at the low multipoles. It is realized from the Fig.\ref{fig:angular} that, the hybrid approach for both potentials provides the closest agreement with both the Planck data and the best-fit $\Lambda$CDM curve at low multipoles considered here. However, the resulting angular power spectrum for the Starobinsky potential is slightly below the best-fit $\Lambda$CDM curve. On the other hand, the dressed metric approach shows a visible deviation from the best-fit $\Lambda$CDM curve for $l<4$; however, the deviation is more for the Starobinsky potential. In other words, with the dressed metric approach, the quadratic potential shows more consistency. For the alternative mass function approach, we also have some deviation but smaller than the dressed metric; however, here the deviation is less for the Starobinsky potential. 

Overall, these results indicate that the sensitivity of the angular power spectrum to the choice of perturbation approach, and to the underlying inflaton potential, is confined to the lowest multipoles, with the hybrid approach giving the best agreement with the Planck data among the three approaches considered here.

\section{Conclusions} \label{Sec5}

In LQC, the big bang singularity is replaced by a quantum bounce, and this has motivated considerable interest in finding whether the bounce and the associated quantum effects can leave any imprint on the observable CMB. In this work, we investigated inflation and the pre-inflationary dynamics in the LQC framework for the quadratic and the Starobinsky potentials, treating the mass of the inflaton as a free parameter. We examined both the background dynamics across the bounce and the resulting evolution of the primordial cosmological perturbations.

The background evolution was explored numerically across the bounce, in both the contracting and the expanding phases, for different values of the mass. Right after the bounce, there is a short super-inflation phase, whose duration was found to be independent of the choice of the potential or of the initial conditions in the KED case. The duration of the damping phase and the total number of e-folds, however, depend on both the shape of the potential and the mass of the field. For the quadratic potential, it was found that when the mass is small, sufficient e-folds can be generated over a wider range of the initial value of $F_B$, whereas for larger mass the acceptable range of $F_B$ becomes narrower. With the quadratic potential, a successful inflation can be produced for both the KED and the PED cases. For the Starobinsky potential, however, this only happens for the KED case; the PED case does not give a suitable picture of inflation with this potential. For the KED case, a smaller value of the initial $F_B$ is required to reach the minimum number of e-folds as the mass of the field decreases. The probability of obtaining the desired slow-roll inflation nonetheless remains very close to $1$ over the whole range of mass considered. It drops slightly with increasing mass; for example, for the quadratic potential, the probability drops from $1$ to about $0.997$, as the mass increases from $10^{-7}m_{Pl}$ to $10^{-3}m_{Pl}$.

Using the background solution, we then computed the primordial power spectrum using three approaches in LQC, the dressed metric, the hybrid, and the alternative mass function approaches. For both potentials and for all the mass values studied, the power spectrum follows the same pattern in every approach. It is suppressed at small $k$, amplified and oscillating over an intermediate range, and becomes nearly scale-invariant at large $k$. It was found that, by increasing the mass, the magnitude of the scale-invariant regime is enhanced and becomes further away from the observed CMB value. We therefore restrict our detailed analysis to a mass of the order of $10^{-6}$ for both potentials. With this mass value, the main difference between the resulting power spectra from the three approaches lies in how fast the power spectrum reaches this scale-invariant regime. The numerical results show that the hybrid approach converges the fastest, followed by the alternative mass function approach, and then the dressed metric approach; this ordering holds for both potentials. We also determined the pivot scale $k_\star$, and the results indicate that it is highly sensitive to the value of the mass, so that for the quadratic potential, a change in mass of about $9\%$ is enough to almost double $k_\star$, while for the Starobinsky potential a change of only about $2\%$ changes $k_\star$ by an order of magnitude.

Finally, we fed the resulting power spectra into the CAMB code to obtain the angular power spectrum and compared it with the Planck 2018 data and the best-fit $\Lambda$CDM model. Since we constrained the pivot scale to stay in the scale-invariant regime, match with the observed CMB value, and exit the horizon at enough e-folds before the end of inflation, the predicted angular power spectrum from all three approaches shows a good consistency for multipolse $l > 4$. for low multipolses, this consistency does not exist for all approaches. The hybrid approach gives the closest match and shows a good agreement with the Planck data for both potentials. The dressed metric, on the other hand, shows the largest deviation, in which we observed more deviation for the Starobinsky potential. The alternative mass function approach also deviates, but by a smaller amount. The deviation here is smaller for the Starobinsky potential. This shows that the potential dependence of the low-multipole deviation is itself approach-dependent. 


Overall, these results show that the choice of perturbation approach barely affects the observable predictions of LQC at scale-invariant regime, but it does leave a distinguishable, potential-dependent signature at low multipoles of the CMB. Future CMB observations, with better precision at low multipoles, may help identify which approach best describes the physics of the bounce.







\begin{thebibliography}{99}
%



\bibitem{inflationtheory1} A. H. Guth, \emph{Inflationary universe: A possible solution to the horizon and flatness problems}, \emph{Phys. Rev. D} {\bf 23} (1981) 347.

\bibitem{inflationtheory2} A. Linde, \emph{A new inflationary universe scenario: A possible solution of the horizon, flatness, homogeneity, isotropy and primordial monopole problems}, \emph{Phys. Lett. B} {\bf 108} (1982) 389.

\bibitem{inflationtheory3} A. Albrecht and P. J. Steinhardt, \emph{Cosmology for Grand Unified Theories with Radiatively Induced Symmetry Breaking}, \emph{Phys. Rev. Lett.} {\bf 48} (1982) 1220.

\bibitem{inflationtheory4} A. Linde, \emph{Chaotic inflation}, \emph{Phys. Lett. B} {\bf 129} (1983) 177.

\bibitem{Gonzalez} J. A. V. Gonzalez, L. E. Padilla, and T. Matos, \emph{Inflationary cosmology: from theory to observations}, \emph{Revista Mexicana de F\'isica C} {\bf 17(1)} (2020) 73.

\bibitem{Lyth} D. H. Lyth and A. R. Liddle, \emph{The primordial density perturbation: cosmology, inflation and the origin of structure}, Cambridge University Press, Cambridge, 2009.

\bibitem{Planck2020} Planck Collaboration, Y. Akrami et al., \emph{Planck 2018 results. X. Constraints on inflation}, \emph{Astron. Astrophys.} {\bf 641} (2020) A10.

\bibitem{Planck2018} Planck Collaboration, N. Aghanim et al., \emph{Planck 2018 results. VI. Cosmological parameters}, \emph{Astron. Astrophys.} {\bf 641} (2020) A6.

\bibitem{Calcagni} G. Calcagni, \emph{Classical and quantum cosmology}, Springer, Berlin, 2017.

\bibitem{Brandenberger} R. Brandenberger, \emph{Initial conditions for inflation—A short review}, \emph{Int. J. Mod. Phys. D} {\bf 26(01)} (2017) 1740002.

\bibitem{Bojowald-Book} M. Bojowald, \emph{Quantum Cosmology: A fundamental description of the Universe}, \emph{Lect. Notes. Phys.} {\bf 835} (2011) 1.

\bibitem{AgulloReview} I. Agullo and P. Singh, \emph{Loop quantum cosmology: A brief review}, arXiv:1612.01236 [gr-qc].

\bibitem{BarrauReview} A. Barrau and B. Bolliet, \emph{Some conceptual issues in loop quantum cosmology}, \emph{Int. J. Mod. Phys. D} {\bf 25(08)} (2016) 1642008.

\bibitem{Rovelli-Book} C. Rovelli, \emph{Quantum Gravity}, Cambridge University Press, Cambridge, 2004.

\bibitem{Thiemann-Book} T. Thiemann, \emph{Modern Canonical Quantum General Relativity}, Cambridge University Press, Cambridge, 2007.

\bibitem{Ashtekar-74} A. Ashtekar, T. Pawlowski, and P. Singh, \emph{Quantum Nature of the Big Bang: Improved dynamics}, \emph{Phys. Rev. D} {\bf 74} (2006) 084003.

\bibitem{Corichi} A. Corichi and E. Montoya, \emph{Coherent semiclassical states for loop quantum cosmology}, \emph{Phys. Rev. D} {\bf 84} (2011) 044021.

\bibitem{Rovelli} C. Rovelli and E. Wilson-Ewing, \emph{Why are effective equations of loop quantum cosmology so accurate?}, \emph{Phys. Rev. D} {\bf 90} (2014) 023538.

\bibitem{Diener} P. Diener, B. Gupt, and P. Singh, \emph{Numerical simulations of a loop quantum cosmos: robustness of the quantum bounce and the validity of effective dynamics}, \emph{Class. Quant. Grav.} {\bf 31} (2014) 105015.

\bibitem{Diener-1} P. Diener, B. Gupt, M. Megevand, and P. Singh, \emph{Numerical evolution of squeezed and non-Gaussian states in loop quantum cosmology}, \emph{Class. Quant. Grav.} {\bf 31} (2014) 165006.

\bibitem{Ashtekar} A. Ashtekar, T. Pawlowski, and P. Singh, \emph{Quantum nature of the big bang}, \emph{Phys. Rev. Lett.} {\bf 96} (2006) 141301.

\bibitem{Ashtekar2} A. Ashtekar, T. Pawlowski, and P. Singh, \emph{Quantum nature of the big bang: Improved dynamics}, \emph{Phys. Rev. D} {\bf 73} (2006) 124038.

\bibitem{Singh} P. Singh and A. Toporensky, \emph{Big crunch avoidance in $k=1$ semiclassical loop quantum cosmology}, \emph{Phys. Rev. D} {\bf 69} (2004) 104008.

\bibitem{Vereshchagin} G. V. Vereshchagin, \emph{A qualitative approach to semi-classical loop quantum cosmology}, \emph{J. Cosmol. Astropart. Phys.} {\bf 0407} (2004) 013.

\bibitem{GDate} G. Date and G. M. Hossain, \emph{Genericness of a Big Bounce in Isotropic Loop Quantum Cosmology}, \emph{Phys. Rev. Lett.} {\bf 94} (2005) 011302.

\bibitem{PSingh} P. Singh, \emph{Are loop quantum cosmos never singular?}, \emph{Class. Quant. Grav.} {\bf 26} (2009) 125005.

\bibitem{PSingh2} P. Singh and F. Vidotto, \emph{Exotic singularities and spatially curved Loop Quantum Cosmology}, \emph{Phys. Rev. D} {\bf 83} (2011) 064027.

\bibitem{PSingh3} P. Singh, \emph{Curvature invariants, geodesics, and the strength of singularities in Bianchi-I loop quantum cosmology}, \emph{Phys. Rev. D} {\bf 85} (2012) 104011.

\bibitem{Kauzuharu} K. Bamba, J. de Haro, and S. D. Odintsov, \emph{Future singularities and teleparallelism in loop quantum cosmology}, \emph{J. Cosmol. Astropart. Phys.} {\bf 1302} (2013) 008.

\bibitem{Joe} A. Joe and P. Singh, \emph{Kantowski-Sachs spacetime in loop quantum cosmology: geometric scalars and the viability of quantization prescriptions}, \emph{Class. Quant. Grav.} {\bf 32} (2015) 015009.

\bibitem{Singh-in} P. Singh, K. Vandersloot, and G. V. Vereshchagin, \emph{Nonsingular bouncing universes in loop quantum cosmology}, \emph{Phys. Rev. D} {\bf 74} (2006) 043510.

\bibitem{Corichi2} A. Corichi and A. Karami, \emph{Measure problem in slow roll inflation and loop quantum cosmology}, \emph{Phys. Rev. D} {\bf 83} (2011) 104006.

\bibitem{Sloan} A. Ashtekar and D. Sloan, \emph{Loop quantum cosmology and slow roll inflation}, \emph{Phys. Lett. B} {\bf 694} (2010) 108.

\bibitem{Sloan2} A. Ashtekar and D. Sloan, \emph{Probability of inflation in loop quantum cosmology}, \emph{Gen. Relativ. Gravit.} {\bf 43} (2011) 3619.

\bibitem{Singh-in-2} B. Gupt and P. Singh, \emph{A quantum gravitational inflationary scenario in Bianchi-I spacetime}, \emph{Class. Quantum Grav.} {\bf 30} (2013) 145013.

\bibitem{Linsefors} L. Linsefors and A. Barrau, \emph{Duration of inflation and conditions at the bounce as a prediction of effective isotropic loop quantum cosmology}, \emph{Phys. Rev. D} {\bf 87} (2011) 123509.

\bibitem{Chen} L. Chen and J. Y. Zhu, \emph{Loop quantum cosmology: The horizon problem and the probability of inflation}, \emph{Phys. Rev. D} {\bf 92} (2015) 084063.

\bibitem{Bedic} S. Bedic and G. Vereshchagin, \emph{Probability of inflation in loop quantum cosmology}, \emph{Phys. Rev. D} {\bf 99} (2019) 043512.

\bibitem{BojowaldSignature1} M. Bojowald and J. Mielczarek, \emph{Some implications of signature-change in cosmological models of loop quantum gravity}, \emph{J. Cosmol. Astropart. Phys.} {\bf 08} (2015) 052.

\bibitem{BojowaldSignature2} M. Bojowald, U. Buyukcam, S. Brahma, and F. D'Ambrosio, \emph{Hypersurface-deformation algebroids and effective space-time models}, \emph{Phys. Rev. D} {\bf 94} (2016) 104032.

\bibitem{BojowaldSignature3} M. Bojowald and S. Brahma, \emph{Signature change in loop quantum gravity: General midisuperspace models and dilaton gravity}, \emph{Phys. Rev. D} {\bf 95} (2017) 124014.

\bibitem{Yang} J. S. Yang, Y. Ding, and Y. G. Ma, \emph{Alternative quantization of the Hamiltonian in loop quantum cosmology}, \emph{Phys. Lett. B} {\bf 682} (2009) 1.

\bibitem{Li084029} B. F. Li, P. Singh, and A. Z. Wang, \emph{Towards cosmological dynamics from loop quantum gravity}, \emph{Phys. Rev. D} {\bf 97} (2018) 084029.

\bibitem{Ashtekar-report} A. Ashtekar and P. Singh, \emph{Loop quantum cosmology: a status report}, \emph{Class. Quantum Grav.} {\bf 28} (2011) 213001.


\bibitem{Mielczarek} J. Mielczarek, T. Cailleteau, J. Grain, and A. Barrau, \emph{Inflation in loop quantum cosmology: dynamics and spectrum of gravitational waves}, \emph{Phys. Rev. D} {\bf 81} (2010) 104049.

\bibitem{Zhang} X. Zhang and Y. Ling, \emph{Inflationary universe in loop quantum cosmology}, \emph{J. Cosmol. Astropart. Phys.} {\bf 08} (2007) 012.

\bibitem{BollietMass} B. Bolliet, J. Grain, C. Stahl, et al., \emph{Comparison of primordial tensor power spectra from the deformed algebra and dressed metric approaches in loop quantum cosmology}, \emph{Phys. Rev. D} {\bf 91} (2015) 084035.

\bibitem{Bonga1} B. Bonga and B. Gupt, \emph{Inflation with the Starobinsky potential in loop quantum cosmology}, \emph{Gen. Relativ. Gravit.} {\bf 48} (2016) 71.

\bibitem{Bonga2} B. Bonga and B. Gupt, \emph{Phenomenological investigation of a quantum gravity extension of inflation with the Starobinsky potential}, \emph{Phys. Rev. D} {\bf 93} (2016) 063513.

\bibitem{Barrau} A. Ashtekar and A. Barrau, \emph{Loop quantum cosmology: From pre-inflationary dynamics to observations}, \emph{Class. Quant. Grav.} {\bf 32} (2015) 234001.

\bibitem{powerlaw} M. Shahalam, M. Sharma, Q. Wu, and A. Z. Wang, \emph{Preinflationary dynamics in loop quantum cosmology: Power-law potentials}, \emph{Phys. Rev. D} {\bf 96} (2017) 123533.

\bibitem{zhu-preinflation} T. Zhu, A. Z. Wang, G. Cleaver, and Q. Sheng, \emph{Universal features of quantum bounce in loop quantum cosmology}, \emph{Phys. Lett. B} {\bf 773} (2017) 196.

\bibitem{zhu-preinflation2} T. Zhu, A. Z. Wang, and G. Cleaver, \emph{Pre-inflationary universe in loop quantum cosmology}, \emph{Phys. Rev. D} {\bf 96} (2017) 083520.

\bibitem{Shahalam2} M. Shahalam, M. Sami, and A. Z. Wang, \emph{Preinflationary dynamics of $\alpha$-attractor in loop quantum cosmology}, \emph{Phys. Rev. D} {\bf 98} (2018) 043524.

\bibitem{Universe} M. Shahalam, \emph{Preinflationary dynamics of power-law potential in loop quantum cosmology}, \emph{Universe} {\bf 4(8)} (2018) 87.

\bibitem{Sharma} M. Sharma, M. Shahalam, Q. Wu, and A. Z. Wang, \emph{Preinflationary dynamics in loop quantum cosmology: Monodromy Potential}, \emph{J. Cosmol. Astropart. Phys.} {\bf 1811} (2018) 003.

\bibitem{inflation} A. Bhardwaj, E. J. Copeland, and J. Louko, \emph{Inflation in loop quantum cosmology}, \emph{Phys. Rev. D} {\bf 99} (2019) 063520.

\bibitem{Zhang2} X. Zhang, J. F. Zhang, J. L. Cui, and L. Zhang, \emph{Chaplygin inflation in loop quantum cosmology}, \emph{Mod. Phys. Lett. A} {\bf 24} (2009) 1763.

\bibitem{SenPRD} A. Sen, \emph{Tachyon field in loop quantum cosmology}, \emph{Phys. Rev. D} {\bf 74} (2006) 043501.

\bibitem{Xiong} H. H. Xiong and J. Y. Zhu, \emph{Tachyon field in loop quantum cosmology: inflation and evolution picture}, \emph{Phys. Rev. D} {\bf 75} (2007) 084023.

\bibitem{Xiao} K. Xiao, X. K. He, F. Huang, and J. Y. Zhu, \emph{Phenomenology analysis of duration inflation for tachyon field in loop quantum cosmology}, \emph{Int. J. Mod. Phys. D} {\bf 23(11)} (2007) 1450087.

\bibitem{XiaoTach} K. Xiao, \emph{Tachyonic inflation in loop quantum cosmology}, \emph{Eur. Phys. J. C} {\bf 79(12)} (2019) 1019.

\bibitem{Mohammadi_DBI} A. Mohammadi, \emph{Exploring the pre-inflationary dynamics in loop quantum cosmology with a DBI scalar field}, \emph{JCAP} {\bf 10} (2024) 062.

\bibitem{MaruganHA1} M. Fernandez-Mendez, G. A. M. Marugan, and J. Olmedo, \emph{Hybrid quantization of an inflationary universe}, \emph{Phys. Rev. D} {\bf 86} (2012) 024003.

\bibitem{MaruganHA2} M. Fernandez-Mendez, G. A. M. Marugan, and J. Olmedo, \emph{Hybrid quantization of an inflationary model: The flat case}, \emph{Phys. Rev. D} {\bf 88} (2013) 044013.

\bibitem{MaruganHA3} L. C. Gomar, M. Fernandez-Mendez, G. A. M. Marugan, and J. Olmedo, \emph{Cosmological perturbations in hybrid loop quantum cosmology: Mukhanov-Sasaki variables}, \emph{Phys. Rev. D} {\bf 90} (2014) 064015.

\bibitem{MaruganHA4} L. C. Gomar, M. Martín-Benito, and G. A. M. Marugan, \emph{Quantum corrections to the Mukhanov-Sasaki equations}, \emph{Phys. Rev. D} {\bf 93} (2016) 104025.

\bibitem{AgulloDM1} I. Agullo, A. Ashtekar, and W. Nelson, \emph{The pre-inflationary dynamics of loop quantum cosmology: confronting quantum gravity with observations}, \emph{Class. Quant. Grav.} {\bf 30} (2013) 085014.

\bibitem{AgulloDM2} I. Agullo, A. Ashtekar, and W. Nelson, \emph{Quantum Gravity Extension of the Inflationary Scenario}, \emph{Phys. Rev. Lett.} {\bf 109} (2012) 251301.

\bibitem{AgulloDM3} I. Agullo, A. Ashtekar, and W. Nelson, \emph{Extension of the quantum theory of cosmological perturbations to the Planck era}, \emph{Phys. Rev. D} {\bf 87} (2013) 043507.

\bibitem{BojowaldCD1} M. Bojowald, G. M. Hossain, M. Kagan, and S. Shankaranarayanan, \emph{Anomaly freedom in perturbative loop quantum gravity}, \emph{Phys. Rev. D} {\bf 78} (2008) 063547.

\bibitem{BojowaldCD2} T. Cailleteau, J. Mielczarek, A. Barrau, and J. Grain, \emph{Anomaly-free scalar perturbations with holonomy corrections in loop quantum cosmology}, \emph{Class. Quant. Grav.} {\bf 29} (2012) 095010.

\bibitem{BojowaldCD3} T. Cailleteau, A. Barrau, F. Vidotto, and J. Grain, \emph{Consistency of holonomy-corrected scalar, vector, and tensor perturbations in loop quantum cosmology}, \emph{Phys. Rev. D} {\bf 86} (2012) 087301.

\bibitem{EwingSU1} E. Wilson-Ewing, \emph{Separate universes in loop quantum cosmology: Framework and applications}, \emph{Int. J. Mod. Phys.} {\bf 25} (2016) 1642002.

\bibitem{EwingSU2} E. Wilson-Ewing, \emph{Testing loop quantum cosmology}, \emph{C. R. Phys.} {\bf 18} (2017) 207.

\bibitem{Bojowald} M. Bojowald, \emph{Consistent loop quantum cosmology}, \emph{Class. Quant. Grav.} {\bf 26} (2009) 075020.

\bibitem{BIP} K. A. Meissner, \emph{Black-hole entropy in loop quantum gravity}, \emph{Class. Quant. Grav.} {\bf 21} (2004) 5245.

\bibitem{Bojowald00044} M. Bojowald, \emph{Effective field theory in loop quantum cosmology}, \emph{Universe} {\bf 5} (2019) 44.

\bibitem{zhangPRL} X. D. Zhang and Y. G. Ma, \emph{Extension of Loop Quantum Gravity to $f(R)$ Theories}, \emph{Phys. Rev. Lett.} {\bf 106} (2011) 171301.

\bibitem{Ellis} J. Ellis, D. V. Nanopoulos, and K. A. Olive, \emph{No-scale supergravity realization of the Starobinsky model of inflation}, \emph{Phys. Rev. Lett.} {\bf 111} (2013) 111301.

\bibitem{Asaka} T. Asaka, S. Iso, H. Kawai, et al., \emph{Reinterpretation of the Starobinsky model}, \emph{Prog. Theor. Exp. Phys.} {\bf 2016(12)} (2016) 123E01.

\bibitem{Pallis} C. Pallis and N. Toumbas, \emph{Starobinsky Inflation: From Non-SUSY to SUGRA Realizations}, \emph{Adv. High Energy Phys.} {\bf 2017} (2017) 6759627.

\bibitem{Li066016} B. F. Li, P. Singh, and A. Z. Wang, \emph{Qualitative dynamics and inflationary attractors in loop cosmology}, \emph{Phys. Rev. D} {\bf 98} (2018) 066016.

\bibitem{Luc} J. Luc and J. Mielczarek, \emph{Slow-roll approximation in loop quantum cosmology}, \emph{J. Cosmol. Astropart. Phys.} {\bf 1701} (2017) 045.


\bibitem{Graef} L. L. Graef and R. O. Ramos, \emph{Probability of warm inflation in loop quantum cosmology}, \emph{Phys. Rev. D} {\bf 98} (2018) 023531.

\bibitem{Barrau-perturbation} A. Barrau, P. Jamet, K. Martineau, et al., \emph{Scalar spectra of primordial perturbations in loop quantum cosmology}, \emph{Phys. Rev. D} {\bf 98(8)} (2018) 086003.

\bibitem{Navascues} B. E. Navascués, D. M. de Blas, and G. A. M. Marugan, \emph{Time-dependent mass of cosmological perturbations in the hybrid and dressed metric approaches to loop quantum cosmology}, \emph{Phys. Rev. D} {\bf 97(4)} (2018) 043523.

\bibitem{ElizagaNavascues:2018bgp} B. Elizaga Navascués, D. M. de Blas, and G. A. M. Marugan, \emph{The Vacuum State of Primordial Fluctuations in Hybrid Loop Quantum Cosmology}, \emph{Universe} {\bf 4(10)} (2018) 98.

\bibitem{Agulloperturbation} I. Agullo, \emph{Primordial power spectrum from the Dapor–Liegener model of loop quantum cosmology}, \emph{Gen. Relativ. Gravit.} {\bf 50(7)} (2018) 91.

\bibitem{Gupt} B. Gupt and B. Bonga, \emph{Phenomenology of inflationary scenario in loop quantum cosmology}, \emph{The Fourteenth Marcel Grossmann Meeting} (2017) pp. 4037-4042.

\bibitem{Olmedo} J. Olmedo and E. Alesci, \emph{Power spectrum of primordial perturbations for an emergent universe in quantum reduced loop gravity}, \emph{J. Cosmol. Astropart. Phys.} {\bf 1904} (2019) 030.


\bibitem{LiWangSingh2020DM} B.-F. Li, P. Singh, and A. Z. Wang, \emph{Primordial power spectrum from the dressed metric approach in loop cosmologies}, \emph{Phys. Rev. D} {\bf 101} (2020) 086004.

\bibitem{LiOlmedoSingWang2020Hyb} B.-F. Li, J. Olmedo, P. Singh, and A. Z. Wang, \emph{Primordial scalar power spectrum from the hybrid approach in loop cosmologies}, \emph{Phys. Rev. D} {\bf 102} (2020) 126025.

\bibitem{LiSingh2022} B.-F. Li and P. Singh, \emph{Close relationship between the dressed metric and the hybrid approach to perturbations in effective loop quantum cosmology}, \emph{Phys. Rev. D} {\bf 106} (2022) 086015.

\bibitem{Lietal_constraining} B.-F. Li, M. Motaharfar, and P. Singh, \emph{Constraining regularization ambiguities in loop quantum cosmology via CMB}, \emph{Phys. Rev. D} {\bf 110} (2024) 066005.

\bibitem{LiSingh2024mass} B.-F. Li and P. Singh, \emph{Alternative effective mass functions in the modified Mukhanov-Sasaki equation of loop quantum cosmology}, \emph{Phys. Rev. D} {\bf 109} (2024) 066005.

\bibitem{MohammadiLiZhu2025} A. Mohammadi, B.-F. Li, and T. Zhu, \emph{Primordial angular power spectrum from the alternative mass function in loop quantum cosmology}, \emph{Phys. Rev. D} {\bf 112} (2025) 044065.

\bibitem{Langlois1994} D. Langlois, \emph{Hamiltonian formalism and gauge invariance for linear perturbations in inflation}, \emph{Class. Quant. Grav.} {\bf 11} (1994) 389.

\bibitem{liddle_2003} A. R. Liddle and S. M. Leach, \emph{How long before the end of inflation were observable perturbations produced?}, \emph{Phys. Rev. D} {\bf 68} (2003) 103503.

\bibitem{Baumann_Tasi} D. Baumann, \emph{TASI Lectures on Inflation}, arXiv:0907.5424 [hep-th].

\bibitem{CAMB} A. Lewis, A. Challinor, and A. Lasenby, \emph{Efficient computation of CMB anisotropies in closed FRW models}, \emph{Astrophys. J.} {\bf 538} (2000) 473.

\end{thebibliography}

\end{document}